\documentclass[english,manuscript,A4,english, APS,  PRD, nofootinbib,superscriptaddress]{revtex4-1}
\usepackage[T1]{fontenc}
\usepackage[latin9]{inputenc}
\usepackage{xcolor}
\usepackage{babel}
\usepackage{amsmath}
\usepackage{graphicx}
\usepackage[unicode=true,pdfusetitle,
 bookmarks=true,bookmarksnumbered=false,bookmarksopen=false,
 breaklinks=false,pdfborder={0 0 0},pdfborderstyle={},backref=false,colorlinks=true]
 {hyperref}
\hypersetup{
 linkcolor=blue, citecolor=cyan}

\makeatletter
\usepackage{amsthm}

\usepackage{xcolor}

\makeatother

\begin{document}
\title{Causality and stability analysis for the minimal causal spin hydrodynamics}
\author{Xin-Qing Xie}
\affiliation{Department of Modern Physics, University of Science and Technology
of China, Anhui 230026, China}
\author{Dong-Lin Wang}
\email{donglinwang@mail.ustc.edu.cn}

\affiliation{Department of Modern Physics, University of Science and Technology
of China, Anhui 230026, China}
\author{Chen Yang}
\affiliation{Department of Modern Physics, University of Science and Technology
of China, Anhui 230026, China}
\author{Shi Pu}
\email{shipu@ustc.edu.cn}

\affiliation{Department of Modern Physics, University of Science and Technology
of China, Anhui 230026, China}
\affiliation{Southern Center for Nuclear-Science Theory (SCNT), Institute of Modern
Physics, Chinese Academy of Sciences, Huizhou 516000, Guangdong Province,
China}
\begin{abstract}
We perform the linear analysis of causality and stability for a minimal
extended spin hydrodynamics up to second order of the gradient expansion.
The first order spin hydrodynamics, with a rank-$3$ spin tensor being
antisymmetric for only the last two indices, are proved to be acausal
and unstable. We then consider the minimal causal spin hydrodynamics
up to second order of the gradient expansion. We derive the necessary
causality and stability conditions for this minimal causal spin hydrodynamics.
Interestingly, the satisfaction of the stability conditions relies
on the equations of state for the spin density and chemical potentials.
Moreover, different with the conventional relativistic dissipative
hydrodynamics, the stability of the theory seems to be broken at the
finite wave-vector when the stability conditions are fulfilled at
small and large wave-vector limits. It implies that the behavior in
small and large wave-vector limits may be insufficient to determine
the stability conditions for spin hydrodynamics in linear mode analysis.
\end{abstract}
\maketitle

\section{Introduction}

%Global polarization 

Relativistic heavy ion collisions provide a novel platform to study
the spin physics. In non-central relativistic heavy-ion collisions,
the quark-gluon plasma (QGP) with large angular momentum perpendicular
to the reaction plane is created. Because of the total angular momentum
conservation, the averaged spin of final particles produced from QGP
are polarized along the direction of the initial orbital angular momentum
\citep{Liang:2004ph,Liang:2004xn,Gao:2007bc}, as known as the global
polarization. The measurements of the global polarization for $\Lambda,\overline{\Lambda}$,
and other hyperons \citep{STAR:2017ckg,STAR:2018gyt,STAR:2019erd,ALICE:2019onw,STAR:2020xbm,Kornas:2020qzi,STAR:2021beb}
can be understood well by various phenomenological models \citep{Becattini:2007nd,Becattini:2007sr,Becattini:2013fla,Fang:2016vpj,Karpenko:2016jyx,Xie:2017upb,Li:2017slc,Sun:2017xhx,Shi:2017wpk,Wei:2018zfb,Xia:2018tes,Vitiuk:2019rfv,Fu:2020oxj,Ryu:2021lnx,Lei:2021mvp,Wu:2022mkr}.
The experimental data also indicates that the QGP generated in non-central
relativistic heavy-ion collisions is the most vortical fluid ever
observed \citep{STAR:2017ckg}. STAR \citep{Niida:2018hfw,STAR:2019erd}
and ALICE Collaboration \citep{ALICE:2021pzu} also measured the local
polarization of $\Lambda$ and $\overline{\Lambda}$ along the beam
and out-of-plane directions. Interestingly, the sign of local polarization
in theoretical calculations is opposite to that of experimental data
\citep{Karpenko:2016jyx,Xie:2017upb,Becattini:2017gcx,Florkowski:2019voj,Wu:2019eyi,Fu:2020oxj}.
To resolve the disagreement, a great deal of effort has been taken
in feed-down effects \citep{Xia:2019fjf,Becattini:2019ntv}, hadronic
interactions \citep{Csernai:2018yok,Nogueira-Santos:2020aky}, relativistic
spin hydrodynamics \citep{Montenegro:2017rbu,Montenegro:2017lvf,Florkowski:2017dyn,Florkowski:2017ruc,Florkowski:2018myy,Becattini:2018duy,Yang:2018lew,Florkowski:2018fap,Florkowski:2018ahw,Hattori:2019lfp,Florkowski:2019qdp,Florkowski:2019voj,Fukushima:2020qta,Fukushima:2020ucl,Shi:2020htn,Li:2020eon,Bhadury:2020puc,Gallegos:2020otk,Garbiso:2020puw,Shi:2020qrx,Bhadury:2020cop,Singh:2020rht,Gallegos:2021bzp,She:2021lhe,Hongo:2021ona,Liu:2020ymh,Florkowski:2021wvk,Wang:2021ngp,Peng:2021ago,Wang:2021wqq,Weickgenannt:2022zxs,Weickgenannt:2022jes,Cao:2022aku,Gallegos:2022jow,Weickgenannt:2022qvh,Daher:2022wzf,Sarwar:2022yzs,Biswas:2022bht,Shi:2023sxh,Biswas:2023qsw},
statistical models \citep{Becattini:2017gcx,Becattini:2022zvf,Palermo:2023cup},
quantum kinetic theory \citep{Stephanov:2012ki,Son:2012zy,Gao:2012ix,Chen:2012ca,Manuel:2013zaa,Manuel:2014dza,Chen:2014cla,Chen:2015gta,Hidaka:2016yjf,Hidaka:2017auj,Mueller:2017lzw,Hidaka:2018ekt,Hidaka:2018mel,Huang:2018wdl,Liu:2018xip,Gao:2018wmr,Gao:2019znl,Weickgenannt:2019dks,Hattori:2019ahi,Wang:2019moi,Li:2019qkf,Lin:2019ytz,Lin:2019fqo,Weickgenannt:2020sit,Weickgenannt:2020aaf,Yang:2020hri,Liu:2020flb,Yamamoto:2020zrs,Gao:2020vbh,Huang:2020wrr,Gao:2020pfu,Wang:2020dws,Weickgenannt:2021cuo,Wang:2021qnt,Sheng:2021kfc,Liu:2021nyg,Hidaka:2022dmn,Fang:2022ttm},
effective theories \citep{Zhang:2019xya,Liu:2020dxg,Liu:2021uhn},
and other phenomenological models \citep{Xie:2017upb,Voloshin:2017kqp,Xia:2018tes,Wei:2018zfb,Liu:2019krs,Xia:2019fjf,Wu:2019eyi,Becattini:2019ntv,Wu:2020yiz,Fu:2020oxj,Florkowski:2021xvy,Sun:2021nsg}.
Although there are much important progress \citep{Liu:2021uhn,Fu:2021pok,Becattini:2021suc,Becattini:2021iol,Yi:2021ryh,Sun:2021nsg,Yi:2021unq,Florkowski:2021xvy,Alzhrani:2022dpi,Wu:2022mkr,Yi:2023tgg},
the local polarization has not been fully understood. Another important
phenomenon related to spin, called the spin alignment of vector mesons
proposed by Refs. \citep{Liang:2004ph,Liang:2004xn,Gao:2007bc}, has
drawn a lot of attentions. The spin alignment is characterized by
the deviation of $\rho_{00}$ from $1/3$, where $\rho_{00}$ is the
$00$ component of the spin density matrix of vector mesons \citep{Schilling:1969um}.
A non-vanishing $\rho_{00}-1/3$ indicates a net spin alignment of
vector mesons. The experimental results \citep{ALICE:2019aid,Mohanty:2021vbt,ALICE:2022sli,STAR:2022fan,Singha:2022syo,Chen:2023hnb}
show that the magnitude of the spin alignment of vector meson is much
larger than that caused by vorticity and other conventional effects
\citep{Liang:2004xn,Yang:2017sdk,Xia:2020tyd,Gao:2021rom,Muller:2021hpe,Kumar:2023ghs}.
Such unexpectedly large spin alignment may arise from a fluctuating
strong force field of $\phi$ \citep{Sheng:2019kmk,Sheng:2020ghv,Sheng:2022wsy,Sheng:2022ffb,Sheng:2023urn}.

%2023.06.15

%Spin hydrodynamics 

The above novel phenomena related to spin triggered the rapid developments
of spin hydrodynamics \citep{Montenegro:2017rbu,Montenegro:2017lvf,Florkowski:2017dyn,Florkowski:2017ruc,Florkowski:2018myy,Becattini:2018duy,Yang:2018lew,Florkowski:2018fap,Florkowski:2018ahw,Hattori:2019lfp,Florkowski:2019qdp,Florkowski:2019voj,Fukushima:2020qta,Fukushima:2020ucl,Shi:2020htn,Li:2020eon,Bhadury:2020puc,Gallegos:2020otk,Garbiso:2020puw,Shi:2020qrx,Bhadury:2020cop,Singh:2020rht,Gallegos:2021bzp,She:2021lhe,Hongo:2021ona,Liu:2020ymh,Florkowski:2021wvk,Wang:2021ngp,Peng:2021ago,Wang:2021wqq,Weickgenannt:2022zxs,Weickgenannt:2022jes,Cao:2022aku,Gallegos:2022jow,Weickgenannt:2022qvh,Daher:2022wzf,Sarwar:2022yzs,Biswas:2022bht,Biswas:2023qsw}.
The spin hydrodynamics is a natural extension of the conventional
hydrodynamics coupled with the dynamic evolution of spin through the
total angular momentum conservation. The spin hydrodynamics incorporating
the quantum property may serve as a powerful tool to understand the
novel phenomena about spin in non-central relativistic heavy-ion collisions.
Over the past few years, various approaches have been proposed to
construct spin hydrodynamics, such as entropy current analysis \citep{Hattori:2019lfp,Fukushima:2020ucl,Li:2020eon,Gallegos:2021bzp,She:2021lhe,Hongo:2021ona,Cao:2022aku,Biswas:2023qsw},
quantum kinetic theory \citep{Florkowski:2017ruc,Florkowski:2017dyn,Hidaka:2017auj,Florkowski:2018myy,Weickgenannt:2019dks,Bhadury:2020puc,Weickgenannt:2020aaf,Shi:2020htn,Speranza:2020ilk,Bhadury:2020cop,Singh:2020rht,Peng:2021ago,Sheng:2021kfc,Weickgenannt:2022zxs,Weickgenannt:2022jes,Weickgenannt:2022qvh},
holographic duality \citep{Gallegos:2020otk,Garbiso:2020puw}, and
effective Lagrangian method \citep{Montenegro:2017rbu,Montenegro:2017lvf}.

%Pseudo-gauge transformations in spin hydrodynamics 

In spite of the substantial efforts, the arbitrariness due to pseudo-gauge
transformations in spin hydrodynamics is not fully understood. Through
the pseudo-gauge transformations \citep{Hehl:1976vr,Leader:2013jra},
one can obtain new forms of energy momentum tensor and spin tensor
without affecting the conservation law. Although such transformations
have no impacts on the total conserved charges, they indeed change
the values of locally defined quantities, e.g., energy momentum tensor
and spin tensor \citep{Hehl:1976vr,Leader:2013jra,Becattini:2018duy,Speranza:2020ilk}.
Thus, different pseudo-gauge transformations give rise to different
frameworks of the spin hydrodynamics, e.g., canonical \citep{Hattori:2019lfp,Hongo:2021ona,Dey:2023hft},
Belinfante \citep{Fukushima:2020ucl}, Hilgevoord-Wouthuysen (HW)
\citep{Hilgevoord:1965,Weickgenannt:2022zxs}, de Groot-van Leeuwen-van
Weert (GLW) \citep{DeGroot:1980dk,Florkowski:2018fap} forms. Which
framework is suitable for understanding the experimental data leads
to intense discussions \citep{Becattini:2018duy,Speranza:2020ilk,Fukushima:2020ucl,Das:2021aar,Buzzegoli:2021wlg,Weickgenannt:2022jes,Daher:2022xon}.

%We consider canonical spin hydrodynamics in this work

So far, the spin hydrodynamics in the first order of the gradient
expansion, with a rank-3 spin tensor that exhibits antisymmetry solely
in its last two indices, has been established \citep{Hattori:2019lfp,Fukushima:2020ucl}.
Before simulating the spin hydrodynamics, it is necessary to investigate
the theory\textquoteright s causality and stability, as is done in
conventional hydrodynamics. In fact, the first order conventional
relativistic hydrodynamics at Landau frame in the gradient expansion
are always acausal and unstable, e.g., see the discussions in Refs.
\citep{Hiscock:1985zz,Hiscock:1987zz,Denicol:2008ha,Pu:2009fj}. Therefore,
the question whether the first order spin hydrodynamics can be casual
or stable arises. Several studies conclude that the spin hydrodynamics
up to the first order in the gradient expansion may be acausal and
unstable in the linear modes analysis \citep{Sarwar:2022yzs,Daher:2022wzf}.
In the early study \citep{Hattori:2019lfp}, the authors have modified
the constitutive relations for the anti-symmetric part of energy momentum
tensor through the equations of motion for the fluid and the stability
conditions of this first order theory in the rest frame of fluid seem
to be satisfied in the linear modes analysis. Later on, the Ref. \citep{Sarwar:2022yzs}
shows that this first order theory may be acausal, while Ref. \citep{Daher:2022wzf}
finds the stability conditions (which corresponds to Eq. (\ref{eq:FirstSixUnstable})
in this work) may not be satisfied.

%In this work  

In this work, we systematically investigate the linear causality and
stability for the spin hydrodynamics proposed in Refs. \citep{Hattori:2019lfp,Fukushima:2020ucl}.
Our findings indicate that the spin hydrodynamics up to the first
order in the gradient expansion is acausal and unstable even when
using the replacement mentioned by Ref. \citep{Hattori:2019lfp}.
The acausal and unstable modes can usually be removed when extending
the theory up to the second order in the gradient expansion. Therefore,
we follow the method outlined in the conventional hydrodynamics \citep{Israel:1979,Israel:1979wp,Koide:2006ef,Denicol:2008ha,Pu:2009fj}
to consider the minimal causal spin hydrodynamics. It is sufficient
to see whether the causality and stability can be recovered up to
the second order in the gradient expansion \citep{Israel:1979,Israel:1979wp,Koide:2006ef,Denicol:2008ha,Pu:2009fj}.
We then analyze the causality and stability for this minimal extended
theory.

The paper is organized as follows. We first review the first order
spin hydrodynamics introduced in Refs. \citep{Hattori:2019lfp,Fukushima:2020ucl}
in Sec. \ref{sec:ReviewFirstOrder} and show it is acausal and unstable
in Sec \ref{sec:CausalityStabilityFirstOrder}. In Sec \ref{subsec:Minimal-causal-theory},
we consider the minimal causal spin hydrodynamics following the method
outlined in the conventional hydrodynamics. In Sec \ref{sec:LinearMinimalCausalSpinHydro},
we analyze the causality and stability for the minimal causal spin
hydrodynamics in the rest frame and comment the results in moving
frames. We summarize this work in Sec. \ref{sec:Conclusion}.

Throughout this work, we work with the metric $g_{\mu\nu}=\mathrm{diag}\{+,-,-,-\}$
and $\Delta_{\mu\nu}=g_{\mu\nu}-u_{\mu}u_{\nu}$. For a rank-$2$
tensor $A^{\mu\nu}$, we introduce the short hand notations $A^{(\mu\nu)}\equiv(A^{\mu\nu}+A^{\nu\mu})/2$,
$A^{[\mu\nu]}\equiv(A^{\mu\nu}-A^{\nu\mu})/2$, and $A^{<\mu\nu>}\equiv\frac{1}{2}[\Delta^{\mu\alpha}\Delta^{\nu\beta}+\Delta^{\mu\beta}\Delta^{\nu\alpha}]A_{\alpha\beta}-\frac{1}{3}\Delta^{\mu\nu}(\Delta^{\alpha\beta}A_{\alpha\beta}).$

\section{First order spin hydrodynamics \label{sec:ReviewFirstOrder}}

In this section, let us briefly review the first order relativistic
spin hydrodynamics. In spin hydrodynamics, we have the conservation
equations for energy, momentum, total angular momentum, and particle
number, i.e., \citep{Hattori:2019lfp,Fukushima:2020ucl,Li:2020eon,Hu:2021lnx,She:2021lhe,Hongo:2021ona,Cao:2022aku}
\begin{equation}
\partial_{\mu}\Theta^{\mu\nu}=0,\quad\partial_{\lambda}J^{\lambda\mu\nu}=0,\quad\partial_{\mu}j^{\mu}=0,\label{eq:ConservationEqs}
\end{equation}
where $\Theta^{\mu\nu}$ is the energy momentum tensor, $J^{\lambda\mu\nu}$
is the total angular momentum current, and $j^{\mu}$ is the current
for particle number. Different with conventional relativistic hydrodynamics,
the total angular momentum conservation equation in Eq. (\ref{eq:ConservationEqs})
plays a crucial role to describe the evolution of spin. The total
angular momentum current can be written as \citep{Hattori:2019lfp,Fukushima:2020ucl}
\begin{eqnarray}
J^{\lambda\mu\nu} & = & x^{\mu}\Theta^{\lambda\nu}-x^{\nu}\Theta^{\lambda\mu}+\Sigma^{\lambda\mu\nu},\label{eq:TotalAngular}
\end{eqnarray}
where the first two terms corresponds to the conventional orbital
angular momentum, and $\Sigma^{\lambda\mu\nu}$ is the rank-$3$ spin
tensor. Using Eq. (\ref{eq:TotalAngular}), the conservation equation
$\partial_{\lambda}J^{\lambda\mu\nu}=0$ can be rewritten as the spin
evolution equation,
\begin{eqnarray}
\partial_{\lambda}\Sigma^{\lambda\mu\nu} & = & -2\Theta^{[\mu\nu]}.\label{eq:SpinEvo}
\end{eqnarray}
Eq. (\ref{eq:SpinEvo}) implies that the anti-symmetric part of energy
momentum tensor $\Theta^{[\mu\nu]}$ is the source for spin, and the
spin can be viewed as a conserved quantity if and only if $\Theta^{[\mu\nu]}=0$.

After introducing the spin degrees of freedom, the thermodynamic relations
in spin hydrodynamics are modified as \citep{Hattori:2019lfp,Fukushima:2020ucl,Li:2020eon,Hu:2021lnx,She:2021lhe,Hongo:2021ona,Cao:2022aku}
\begin{eqnarray}
e+p & = & Ts+\mu n+\omega_{\mu\nu}S^{\mu\nu},\label{eq:ThermalR1}\\
de & = & Tds+\mu dn+\omega_{\mu\nu}dS^{\mu\nu}.\label{eq:ThermalR2}
\end{eqnarray}
where $e,p,T,s,n,\mu,\omega_{\mu\nu}$, and $S^{\mu\nu}$ denote energy
density, pressure, temperature, entropy density, particle number density,
chemical potential, spin chemical potential, and spin density. The
spin density is defined as 
\begin{equation}
S^{\mu\nu}\equiv u_{\lambda}\Sigma^{\lambda\mu\nu}\label{eq:DefinitionOfSpinDensity}
\end{equation}
with the fluid velocity $u^{\mu}$. Analogy to the relationship between
$\mu$ and $n$, here we introduce the anti-symmetric spin chemical
potential $\omega_{\mu\nu}$ as the conjugate of $S^{\mu\nu}$. 

Before decomposing the $\Theta^{\mu\nu}$ and $\Sigma^{\lambda\mu\nu}$,
we emphasize that there exist different choices for them. For example,
by applying the N\"other theorem to two equivalent Lagrangian density
for Dirac field,
\begin{eqnarray}
\mathcal{L}_{1} & = & \overline{\psi}(i\gamma\cdot\partial-m)\psi,\\
\mathcal{L}_{2} & = & \frac{1}{2}\overline{\psi}i\gamma\cdot\overset{\leftrightarrow}{\partial}_{\mu}\psi-m\overline{\psi}\psi,
\end{eqnarray}
where $\overset{\leftrightarrow}{\partial}_{\mu}\equiv\overset{\rightarrow}{\partial}_{\mu}-\overset{\leftarrow}{\partial}_{\mu}$,
two distinct sets of energy momentum tensors and spin tensors emerge,
\begin{eqnarray}
\Theta_{1}^{\mu\nu} & = & \overline{\psi}i\gamma^{\mu}\partial^{\nu}\psi,\quad\Sigma_{1}^{\lambda\mu\nu}=\frac{1}{4}\overline{\psi}i\gamma^{\lambda}[\gamma^{\mu},\gamma^{\nu}]\psi,\\
\Theta_{2}^{\mu\nu} & = & \frac{i}{2}\overline{\psi}\gamma^{\mu}\overset{\leftrightarrow}{\partial^{\nu}}\psi,\quad\Sigma_{2}^{\lambda\mu\nu}=\frac{1}{8}\overline{\psi}i\{\gamma^{\lambda},[\gamma^{\mu},\gamma^{\nu}]\}\psi.
\end{eqnarray}
Here, $\Sigma_{1}^{\lambda\mu\nu}$ is antisymmetric only with respect
to $\mu$ and $\nu$ indices, while $\Sigma_{2}^{\lambda\mu\nu}$
is totally antisymmetric. In principle, one can derive the spin hydrodynamics
from the microscopic theories, as demonstrated in Refs. \citep{Florkowski:2018fap,Bhadury:2022ulr,Weickgenannt:2022zxs,Weickgenannt:2022jes,Weickgenannt:2022qvh}
for kinetic theories and Refs. \citep{Becattini:2012tc,Becattini:2018duy,Becattini:2020sww}
for statistical methods. An alternative method to derive the spin
hydrodynamics is to map the tensor structure of hydrodynamic variables
to operators mentioned above, e.g. see Refs. \citep{Hattori:2019lfp,Fukushima:2020ucl,Hongo:2021ona}.
In this work, we follow Refs. \citep{Hattori:2019lfp,Fukushima:2020ucl}
and adopt the energy momentum tensor and spin tensor sharing the similar
tensor structure with $\Theta_{1}^{\mu\nu}$ and $\Sigma_{1}^{\lambda\mu\nu}$,
respectively. For other choices, one can refer to Refs. \citep{Speranza:2020ilk,Hongo:2021ona,Daher:2022xon}
and references therein. 

Following Refs. \citep{Hattori:2019lfp,Fukushima:2020ucl}, the energy
momentum tensor and particle current can be decomposed as 
\begin{eqnarray}
\Theta^{\mu\nu} & = & eu^{\mu}u^{\nu}-(p+\Pi)\Delta^{\mu\nu}+2h^{(\mu}u^{\nu)}+\pi^{\mu\nu}+2q^{[\mu}u^{\nu]}+\phi^{\mu\nu},\label{eq:Thetamunu}\\
j^{\mu} & = & nu^{\mu}+\nu^{\mu},
\end{eqnarray}
where $h^{\mu},\nu^{\mu}$, $\Pi$, and $\pi^{\mu\nu}$ stand for
heat current, particle diffusion, bulk viscous pressure, and shear
stress tensor, respectively, and the antisymmetric parts $2q^{[\mu}u^{\nu]}$
and $\phi^{\mu\nu}$ are related to the spin effects. As for the rank-3
spin tensor $\Sigma^{\lambda\mu\nu}$, we have \citep{Hattori:2019lfp,Fukushima:2020ucl}
\begin{eqnarray}
\Sigma^{\lambda\mu\nu} & = & u^{\lambda}S^{\mu\nu}+\Sigma_{(1)}^{\lambda\mu\nu},\label{eq:spindof6}
\end{eqnarray}
where the spin density $S^{\mu\nu}$ defined in Eq. (\ref{eq:DefinitionOfSpinDensity})
has six independent degrees of freedom.

In this work, we follow the power counting scheme in Refs. \citep{Fukushima:2020ucl,Wang:2021ngp,Wang:2021wqq},
\begin{equation}
S^{\mu\nu}\sim O(1),\ \omega_{\mu\nu}\sim O(\partial),\ \Sigma_{(1)}^{\lambda\mu\nu}\sim O(\partial).\label{eq:PowerCounting-1}
\end{equation}
The spin density $S^{\mu\nu}$ is chosen as the leading order in the
gradient expansion. It corresponds to the case in which the most of
particles in the system are polarized, i.e. the order of $S^{\mu\nu}$
is considered as the same as the one for the number density $n$.
While in Refs. \citep{Hattori:2019lfp,Hongo:2021ona}, the authors
have chosen a different power counting scheme, $S^{\mu\nu}\sim O(\partial)$,
$\omega_{\mu\nu}\sim O(\partial)$, $\Sigma_{(1)}^{\lambda\mu\nu}\sim O(\partial^{2})$.

Following \citep{Hattori:2019lfp,Fukushima:2020ucl}, it is straightforward
to get the entropy production rate, 
\begin{eqnarray}
\partial_{\mu}\mathcal{S}_{\textrm{can}}^{\mu} & = & (h^{\mu}-\frac{e+p}{n}\nu^{\mu})\left[\partial_{\mu}\frac{1}{T}+\frac{1}{T}(u\cdot\partial)u_{\mu}\right]+\frac{1}{T}\pi^{\mu\nu}\partial_{\mu}u_{\nu}-\frac{1}{T}\Pi(\partial\cdot u)\nonumber \\
 &  & +\frac{1}{T}\phi^{\mu\nu}(\partial_{\mu}u_{\nu}+2\omega_{\mu\nu})+\frac{q^{\mu}}{T}\left[T\partial_{\mu}\frac{1}{T}-(u\cdot\partial)u_{\mu}+4\omega_{\mu\nu}u^{\nu}\right]+O(\partial^{3}),\label{eq:entropy_flow_01}
\end{eqnarray}
where $\mathcal{S}_{\textrm{can}}^{\mu}$ is the entropy density current.
The second law of thermodynamics $\partial_{\mu}\mathcal{S}_{\textrm{can}}^{\mu}\geq0$
can give us the first order constitutive relations \citep{Hattori:2019lfp,Fukushima:2020ucl},
\begin{eqnarray}
h^{\mu}-\frac{e+p}{n}\nu^{\mu} & = & \kappa\Delta^{\mu\nu}\left[\frac{1}{T}\partial_{\nu}T-(u\cdot\partial)u_{\nu}\right],\label{eq:SixHeatCurrent}\\
\pi^{\mu\nu} & = & 2\eta\partial^{<\mu}u^{\nu>},\label{eq:SixTau}\\
\Pi & = & -\zeta\partial_{\mu}u^{\mu},\label{eq:SixPi}\\
q^{\mu} & = & \lambda\Delta^{\mu\nu}\left[\frac{1}{T}\partial_{\nu}T+(u\cdot\partial)u_{\nu}-4\omega_{\nu\alpha}u^{\alpha}\right],\label{eq:SixQ}\\
\phi^{\mu\nu} & = & 2\gamma_{s}\Delta^{\mu\rho}\Delta^{\nu\sigma}(\partial_{[\rho}u_{\sigma]}+2\omega_{\rho\sigma}),\label{eq:SixPhi}
\end{eqnarray}
where the heat conductivity coefficient $\kappa$, shear viscosity
coefficient $\eta$, and bulk viscosity $\zeta$ also exist in conventional
hydrodynamics, while $\lambda$ and $\gamma_{s}$ are new coefficients
corresponding to the interchange of spin and orbital angular momentum.
The entropy principle also requires that the transport coefficients
\begin{equation}
\kappa,\eta,\zeta,\lambda,\gamma_{s}>0,\label{eq:coefficients_01}
\end{equation}
are positive. As the system approaches global equilibrium, the entropy
production rate in Eq.(\ref{eq:entropy_flow_01}) tends to zero. It
yields the well-known killing condition \citep{Becattini:2012tc,Becattini:2018duy},
which causes the right hand sides of Eqs. (\ref{eq:SixHeatCurrent},\ref{eq:SixTau},\ref{eq:SixPi},\ref{eq:SixQ},\ref{eq:SixPhi})
to vanish. Especially, we have $q^{\mu},\phi^{\mu\nu}=0$ such that
the energy momentum tensor $\Theta^{\mu\nu}$ is symmetric in global
equilibrium state. Note that, pointed out by Refs. \citep{Cao:2022aku,Hu:2022azy},
some cross terms between the different dissipative currents may also
exist due to the Onsager relation, but here we neglect them for simplicity.

Before ending this section, we would like to comment on the heat flow
$h^{\mu}$. Interestingly, when we set $\nu^{\mu}=0$ and $n=0$,
we find that one cannot fix the expression for heat current $h^{\mu}$
in the first order of gradient expansion. By using $\Delta_{\nu\alpha}\partial_{\mu}\Theta^{\mu\nu}=0$
and Eqs. (\ref{eq:ThermalR1},\ref{eq:ThermalR2}), we find that $[\partial_{\mu}\frac{1}{T}+\frac{1}{T}(u\cdot\partial)u_{\mu}]\sim O(\partial^{2})$
when $\nu^{\mu}=0$ and $n=0$. In that case, the term $h^{\mu}[\partial_{\mu}\frac{1}{T}+\frac{1}{T}(u\cdot\partial)u_{\mu}]\sim O(\partial^{3}),$
will be neglected in the entropy production rate (\ref{eq:entropy_flow_01}),
i.e., we cannot determine the expression of $h^{\mu}$ by the entropy
principle there. The similar behavior was also observed in conventional
hydrodynamics \citep{Kovtun:2019hdm,Bemfica:2020zjp}.

\section{Unstable and acausal modes in the first order spin hydrodynamics
\label{sec:CausalityStabilityFirstOrder}}

In this section, we analyze the causality and stability for the first
order spin hydrodynamics. It is well-known that the conventional relativistic
hydrodynamics in Landau frame up to the first order in gradient expansion
are always acausal, e.g., see Refs. \citep{Hiscock:1985zz,Hiscock:1987zz}
as the early pioneer works.

In linear modes analysis, one can consider the perturbations $\delta X$
to the hydrodynamical quantities $X$ in the equilibrium. By assuming
the $\delta X\sim\delta\tilde{X}e^{i\omega t-ikx}$ with $\delta\tilde{X}$
being constant in space-time, one can solve the dispersion relation
$\omega=\omega(k)$ from the conservation equations. In the conventional
hydrodynamics, the causality condition is usually given by \citep{Pu:2009fj,Kovtun:2019hdm,Hoult:2020eho,Brito:2020nou,Brito:2021iqr}
\begin{equation}
\lim_{k\rightarrow\infty}\left|\textrm{Re }\frac{\omega}{k}\right|\leq1,\label{eq:casuality_01}
\end{equation}
where the condition (\ref{eq:casuality_01}) can also be written as
$\lim_{k\rightarrow\infty}\left|\textrm{Re }\frac{\partial\omega}{\partial k}\right|\leq1$
in some literature \citep{Pu:2009fj,Brito:2020nou,Brito:2021iqr}.
However, the above condition is insufficient to guarantee the causality.
We need an extra condition that, \citep{Krotscheck1978CausalityC}
\begin{equation}
\lim_{k\rightarrow\infty}\left|\frac{\omega}{k}\right|\textrm{ is bounded.}\label{eq:causality_02}
\end{equation}
As pointed out by the early pioneer work \citep{Krotscheck1978CausalityC},
the unbounded $\lim_{k\rightarrow\infty}\left|\frac{\omega}{k}\right|$
gives the infinite propagating speed of the perturbation, even if
the $\omega$ is pure imaginary. One simple example is the non-relativistic
diffusion equation, $\partial_{t}n-D_{n}\partial_{x}^{2}n=0$ with
$D_{n}$ being the diffusion constant. It is easy to check that its
dispersion relation gives $\omega=iD_{n}k^{2}$, which satisfies condition
(\ref{eq:casuality_01}) but does not obey condition (\ref{eq:causality_02}).
Therefore, the perturbation in the non-relativistic diffusion equations
has the unlimited propagating speed, i.e. with any compact initial
value for $n(t_{0},x)$, the $n(t_{0}+\Delta t,x)$ at $x\rightarrow\infty$
can still get the influence \citep{Evans2010PartialDE}. We emphasize
that the conditions (\ref{eq:casuality_01},\ref{eq:causality_02})
are necessary but not sufficient to guarantee that the theory is casual
\citep{Gavassino:2023mad,Wang:2023csj,Hoult:2023clg}. One example
is the transverse perturbations of an Eckart fluid with shear viscous
tensor, whose dispersion relation satisfies the conditions (\ref{eq:casuality_01},\ref{eq:causality_02}),
but the velocity can exceed the speed of light (see Eqs. (47) and
(48) in Ref. \citep{Hiscock:1987zz} for the perturbation equations
and the propagating velocity). 

The stability means that the the imaginary part of $\omega=\omega(k)$
must be positive for $k\neq0$, i.e. 
\begin{equation}
\textrm{Im }\omega(k)>0.\label{eq:Stablity_condition_01}
\end{equation}
Note that the case of $\textrm{Im }\omega=0$ corresponds to the neutral
equilibrium, which means the equilibrium state is not unique. In this
work, we will not consider such special cases, and we only consider
the condition (\ref{eq:Stablity_condition_01}) to study the stability
of spin hydrodynamics as in Ref. \citep{Daher:2022wzf}.

It is necessary to study the causality and stability for the relativistic
spin hydrodynamics in the first order. To see whether the first order
spin hydrodynamics can be casual or not, we consider the linear modes
analysis to the system, i.e. we take the small perturbations on top
of static equilibrium. Following Refs. \citep{Hiscock:1985zz,Hiscock:1987zz},
the static equilibrium background is assumed to be irrotational global
equilibrium state. We label the quantities with subscript $(0)$ as
those at the global equilibrium state, while we use \textquotedblleft $\delta X$\textquotedblright{}
to denote the small perturbations of the quantity $X$, e.g., $e_{(0)}$
and $\delta e$ stand for the energy density at the global equilibrium
and the small perturbations of energy density, respectively.

From now on, unless specified otherwise, we adopt the Landau frame,
and neglect the conserved charge current $j^{\mu}$.

We now consider the small perturbations on top of static equilibrium.
Not all of the perturbations are independent of each other, and we
can choose 
\begin{equation}
\delta e,\ \delta u^{i},\ \delta S^{\mu\nu},
\end{equation}
as independent variables.

The variation of pressure $\delta p$ and spin chemical potential
$\delta\omega^{\mu\nu}$ can be expressed as functions of $\delta e$
and $\delta S^{\mu\nu}$ through 
\begin{equation}
\delta p=c_{s}^{2}\delta e,\quad\delta\omega^{0i}=\chi_{b}\delta S^{0i}+\chi_{e}^{0i}\delta e,\quad\delta\omega^{ij}=\chi_{s}\delta S^{ij}+\chi_{e}^{ij}\delta e,\label{eq:Notations0}
\end{equation}
where the speed of sound $c_{s}$, and $\chi_{b}$, $\chi_{s}$,$\chi_{e}^{\mu\nu}$
are in general the functions of thermodynamic variables. For simplicity,
we take $c_{s}$, $\chi_{b}$, $\chi_{s}$,$\chi_{e}^{\mu\nu}$ as
constants in the linear modes analysis. Note that $\chi_{e}^{\mu\nu}$
comes from the anisotropy of the system. Under the assumption of an
irrotational global equilibrium, from Eq. (\ref{eq:SixQ}) the spin
chemical potential vanishes $\omega_{(0)}^{\mu\nu}=0$. For simplicity,
we further choose $S_{(0)}^{\mu\nu}=0$. The variation of the temperature
$\delta T$ can be obtained by the thermodynamics relations, with
the help of Eqs. (\ref{eq:ThermalR1},\ref{eq:ThermalR2}), 
\begin{equation}
\delta T=\frac{T_{(0)}}{e_{(0)}+p_{(0)}}\left[\delta p-T_{(0)}S_{(0)}^{\mu\nu}\delta\left(\frac{\omega_{\mu\nu}}{T}\right)\right]=\frac{T_{(0)}c_{s}^{2}\delta e}{e_{(0)}+p_{(0)}}.\label{eq:deltaT}
\end{equation}

Next, we consider the variation of the conservation equations $\partial_{\mu}\delta\Theta^{\mu\nu}=0$
and $\partial_{\lambda}\delta J^{\lambda\mu\nu}=0$, where the perturbations
$\delta\Theta^{\mu\nu}$ and $\delta J^{\lambda\mu\nu}$ can be derived
from the constitutive relations in Eqs. (\ref{eq:TotalAngular},\ref{eq:Thetamunu},\ref{eq:SixHeatCurrent}-\ref{eq:SixPhi}).
It is straightforward to obtain the linearized equations for the independent
perturbations $\delta e,\delta\vartheta^{i},\delta S^{\mu\nu}$,

\begin{eqnarray}
0 & = & (\partial_{0}+\frac{1}{2}\lambda^{\prime}c_{s}^{2}\partial_{i}\partial^{i}+4\lambda\chi_{e}^{0i}\partial_{i})\delta e+(\partial_{i}+\frac{1}{2}\lambda^{\prime}\partial_{i}\partial_{0})\delta\vartheta^{i}+D_{b}\partial_{i}\delta S^{0i},\label{eq:Six1stPEq-1}\\
0 & = & (4\gamma_{s}\chi_{e}^{ij}\partial_{i}-c_{s}^{2}\partial^{j}-\frac{1}{2}c_{s}^{2}\lambda^{\prime}\partial_{0}\partial^{j}-4\lambda\chi_{e}^{0j}\partial_{0})\delta e+(\gamma_{\|}-\gamma_{\perp}-\gamma^{\prime})\partial^{j}\partial_{i}\delta\vartheta^{i}\nonumber \\
 &  & +[\partial_{0}-\frac{1}{2}\lambda^{\prime}\partial_{0}\partial_{0}+(\gamma_{\perp}+\gamma^{\prime})\partial^{i}\partial_{i}]\delta\vartheta^{j}-D_{b}\partial_{0}\delta S^{0j}+D_{s}\partial_{i}\delta S^{ij},\\
0 & = & (\lambda^{\prime}c_{s}^{2}\partial^{i}+8\lambda\chi_{e}^{0i})\delta e+\lambda^{\prime}\partial_{0}\delta\vartheta^{i}+(2D_{b}-\partial_{0})\delta S^{0i},\\
0 & = & 8\gamma_{s}\chi_{e}^{ij}\delta e+2\gamma^{\prime}\partial^{i}\delta\vartheta^{j}-2\gamma^{\prime}\partial^{j}\delta\vartheta^{i}+(2D_{s}+\partial_{0})\delta S^{ij}.\label{eq:Six1stPEq-4}
\end{eqnarray}
Here we introduce the following shorthand notations, 
\begin{eqnarray}
D_{s} & \equiv & 4\gamma_{s}\chi_{s},\quad D_{b}\equiv4\lambda\chi_{b},\quad\delta\vartheta^{i}\equiv(e_{(0)}+p_{(0)})\delta u^{i},\quad\lambda^{\prime}\equiv\frac{2\lambda}{e_{(0)}+p_{(0)}},\nonumber \\
\gamma^{\prime} & \equiv & \frac{\gamma_{s}}{e_{(0)}+p_{(0)}},\quad\gamma_{\perp}\equiv\frac{\eta}{e_{(0)}+p_{(0)}},\quad\gamma_{\|}\equiv\frac{\frac{4}{3}\eta+\zeta}{e_{(0)}+p_{(0)}}.\label{eq:Notations1}
\end{eqnarray}

In linear modes analysis, the perturbations are assumed along the
$x$ direction only, 
\begin{equation}
\delta e=\delta\tilde{e}e^{i\omega t-ikx},\ \delta\vartheta^{i}=\delta\tilde{\vartheta}^{i}e^{i\omega t-ikx},\ \delta S^{\mu\nu}=\delta\tilde{S}^{\mu\nu}e^{i\omega t-ikx},\label{eq:PlaneWave1}
\end{equation}
where $\delta\tilde{e}$, $\delta\tilde{\vartheta}^{i}$, and $\delta\tilde{S}^{\mu\nu}$
are independent of space and time.

Inserting the perturbations in Eq. (\ref{eq:PlaneWave1}) into Eqs.
(\ref{eq:Six1stPEq-1}-\ref{eq:Six1stPEq-4}) yields,
\begin{eqnarray}
\mathcal{M}_{1}\delta\tilde{X}_{1} & = & 0,\label{eq:Six1stFourierEq}
\end{eqnarray}
where 
\begin{equation}
\delta\tilde{X}_{1}\equiv(\delta\tilde{e},\delta\tilde{\vartheta}^{x},\delta\tilde{S}^{0x},\delta\tilde{\vartheta}^{y},\delta\tilde{S}^{0y},\delta\tilde{S}^{xy},\delta\tilde{\vartheta}^{z},\delta\tilde{S}^{0z},\delta\tilde{S}^{xz},\delta\tilde{S}^{yz})^{\mathrm{T}},\label{eq:Six1stFourierVariables}
\end{equation}
and 
\begin{equation}
\mathcal{M}_{1}\equiv\left(\begin{array}{cccc}
M_{1} & 0 & 0 & 0\\
A_{1} & M_{2} & 0 & 0\\
A_{2} & 0 & M_{2} & 0\\
A_{3} & 0 & 0 & M_{3}
\end{array}\right),\label{eq:Six1stFourierMatrix}
\end{equation}
with 
\begin{eqnarray}
M_{1} & \equiv & \left(\begin{array}{ccc}
i\omega+\frac{1}{2}\lambda^{\prime}c_{s}^{2}k^{2}-4ik\lambda\chi_{e}^{0x} & \frac{1}{2}\lambda^{\prime}k\omega-ik & -ikD_{b}\\
\frac{1}{2}\lambda^{\prime}c_{s}^{2}k\omega-ikc_{s}^{2}-4i\omega\lambda\chi_{e}^{0x} & \gamma_{\|}k^{2}+i\omega+\frac{1}{2}\lambda^{\prime}\omega^{2} & -i\omega D_{b}\\
ik\lambda^{\prime}c_{s}^{2}+8\lambda\chi_{e}^{0x} & i\omega\lambda^{\prime} & 2D_{b}-i\omega
\end{array}\right),\label{eq:Six1stM1}\\
M_{2} & \equiv & \left(\begin{array}{ccc}
k^{2}(\gamma_{\perp}+\gamma^{\prime})+i\omega+\frac{1}{2}\lambda^{\prime}\omega^{2} & -i\omega D_{b} & -ikD_{s}\\
i\omega\lambda^{\prime} & 2D_{b}-i\omega & 0\\
2ik\gamma^{\prime} & 0 & 2D_{s}+i\omega
\end{array}\right),\label{eq:Six1stM2}\\
M_{3} & \equiv & 2D_{s}+i\omega.\label{eq:Six1stM3}
\end{eqnarray}
The off-diagonal blocks $A_{1}$, $A_{2}$, $A_{3}$ in the matrix
$\mathcal{M}_{1}$, whose expressions are shown in Appendix \ref{sec:Off-diagonal-submatrices-in},
and are irrelevant to the following discussions. The non-trivial solutions
in Eq. (\ref{eq:Six1stFourierEq}) requires,
\begin{equation}
0=\det\mathcal{M}_{1}=\det M_{1}\cdot(\det M_{2})^{2}\cdot\det M_{3}.\label{eq:Six1stDetVanish}
\end{equation}
From Eqs. (\ref{eq:Six1stM1}-\ref{eq:Six1stM3}), we find that Eq.
(\ref{eq:Six1stDetVanish}) is a polynomial equation for two variables
$\omega$ and $k$. Solving this equation gives the dispersion relations
$\omega=\omega(k)$.

The $\det M_{3}=0$ gives a non-hydrodynamic mode,
\begin{eqnarray}
\omega & = & 2iD_{s},\label{eq:Six1stSolM3}
\end{eqnarray}
which corresponds to the spin relaxation \citep{Hattori:2019lfp,Hongo:2021ona}.
The stability condition (\ref{eq:Stablity_condition_01}) requires
that $D_{s}>0$.

The dispersion relation solved from $\det M_{1}=0$ and $\det M_{2}=0$
are lengthy and complicated, so here we only discuss the relations
in small $k$ and large $k$ limits to analyze stability and causality.
In the $k\rightarrow0$ limit, the dispersion relations are 
\begin{eqnarray}
\omega & = & \pm c_{s}k+\frac{i}{2}(\gamma_{\|}\mp4c_{s}\lambda\chi_{e}^{0x}D_{b}^{-1})k^{2}+O(k^{3}),\label{eq:Six1stSolM1Sk-1}\\
\omega & = & (-i\pm\sqrt{4D_{b}\lambda^{\prime}-1})\lambda^{\prime-1}+O(k),\label{eq:Six1stSolM1Sk-2}\\
\omega & = & i\gamma_{\perp}k^{2}+O(k^{3}),\label{eq:Six1stSolM2Sk-1}\\
\omega & = & 2iD_{s}+O(k^{2}).\label{eq:Six1stSolM2Sk-2}
\end{eqnarray}
where the dispersion relations (\ref{eq:Six1stSolM1Sk-1}-\ref{eq:Six1stSolM1Sk-2})
and (\ref{eq:Six1stSolM1Sk-2}-\ref{eq:Six1stSolM2Sk-2}) are solved
from $\det M_{1}=0$ and $\det M_{2}=0$, respectively. The modes
in Eq.  (\ref{eq:Six1stSolM1Sk-1}) and Eq. (\ref{eq:Six1stSolM2Sk-1})
correspond to the sound and shear modes in the conventional hydrodynamics
\citep{Hiscock:1985zz,Denicol:2008ha,Pu:2009fj,Brito:2020nou}, respectively.
The stability condition (\ref{eq:Stablity_condition_01}) for the
dispersion relation in Eqs. (\ref{eq:Six1stSolM1Sk-1}-\ref{eq:Six1stSolM2Sk-2})
gives,
\begin{equation}
D_{s}>0,\ \lambda^{\prime}<0,\ D_{b}<-4c_{s}\lambda\gamma_{\|}^{-1}\big|\chi_{e}^{0x}\big|\leq0.\label{eq:FirstSixUnstable}
\end{equation}
However, conditions (\ref{eq:FirstSixUnstable}) contradict with the
entropy principle in Eq. (\ref{eq:coefficients_01}), i.e. $\lambda^{\prime}=2\lambda/(e_{(0)}+p_{(0)})>0$
defined in Eq. (\ref{eq:Notations1}) with $\lambda>0$ and $e_{(0)}+p_{(0)}>0$.

In the $k\rightarrow\infty$ limit, the dispersion relations become,
\begin{eqnarray}
\omega & = & -4iD_{b}\gamma_{\|}^{-1}\lambda^{\prime-1}k^{-2}+O(k^{-3}),\label{eq:Six1stSolM1Lk-1}\\
\omega & = & -ic_{s}^{2/3}\gamma_{\|}^{1/3}k^{4/3}+O(k),\label{eq:Six1stSolM1Lk-2}\\
\omega & = & (-1)^{1/6}c_{s}^{2/3}\gamma_{\|}^{1/3}k^{4/3}+O(k),\label{eq:Six1stSolM1Lk-3}\\
\omega & = & (-1)^{5/6}c_{s}^{2/3}\gamma_{\|}^{1/3}k^{4/3}+O(k),\label{eq:Six1stSolM1Lk-4}\\
\omega & = & -2iD_{b}+O(k^{-1}),\label{eq:Six1stSolM1Lk-5}\\
\omega & = & 2iD_{s}\gamma_{\perp}(\gamma^{\prime}+\gamma_{\perp})^{-1}+O(k^{-1}),\label{eq:Six1stSolM1Lk-6}\\
\omega & = & \pm ik\sqrt{2\lambda^{\prime-1}(\gamma^{\prime}+\gamma_{\perp})}+O(k^{0}),\label{eq:Six1stSolM1Lk-7}
\end{eqnarray}
where first four modes come from $\det M_{1}=0$, and others can be
derived by $\det M_{2}=0$. Obviously, Eq. (\ref{eq:Six1stSolM1Lk-7})
contains an unstable mode.

On the other hand, we also find that in Eqs. (\ref{eq:Six1stSolM1Lk-2}-\ref{eq:Six1stSolM1Lk-4})
$|\omega/k|$ is unbounded, which violates the causality condition
(\ref{eq:causality_02}). We also notice that Ref. \citep{Sarwar:2022yzs}
has also analyzed the causality for the first order spin hydrodynamics
in small $k$ limit.

We find that the first order spin hydrodynamics is acausal and unstable
similar as the conventional relativistic hydrodynamics in Landau frame.

%2023.06.13

Before ending this section, we comment on the condition (\ref{eq:FirstSixUnstable}).
We notice that the dispersion relations in Refs. \citep{Hattori:2019lfp,Sarwar:2022yzs,Daher:2022wzf}
are different with ours in Eqs. (\ref{eq:Six1stSolM3}-\ref{eq:Six1stSolM1Lk-7}).
Let us explain what happens here. The energy momentum conservation
equation $\Delta_{\mu\alpha}\partial_{\nu}\Theta^{\mu\nu}=0$, gives
the acceleration equations for the fluid velocity, 
\begin{equation}
(u\cdot\partial)u^{\mu}=\frac{1}{T}\Delta^{\mu\nu}\partial_{\nu}T+O(\partial^{2}).\label{eq:EOM_01}
\end{equation}
In Refs.\citep{Hattori:2019lfp,Sarwar:2022yzs,Daher:2022wzf}, the
authors have replaced $(u\cdot\partial)u^{\mu}$ in $q^{\mu}$ in
Eq. (\ref{eq:SixQ}) by Eq.  (\ref{eq:EOM_01}) and gotten another
expression for $q^{\mu}$ 
\begin{equation}
q^{\mu}=\lambda\left(\frac{2\Delta^{\mu\nu}\partial_{\nu}p}{e+p}-4\omega^{\mu\nu}u_{\nu}\right)+O(\partial^{2}).\label{eq:RewriteSixQ}
\end{equation}
Although $q^{\mu}$ in Eq. (\ref{eq:RewriteSixQ}) (also in Refs.
\citep{Hattori:2019lfp,Sarwar:2022yzs,Daher:2022wzf} ) is equivalent
to our $q^{\mu}$ in Eq. (\ref{eq:SixQ}) up to the first order in
gradient expansion, we emphasize that these two $q^{\mu}$ correspond
to different hydrodynamic frames and will lead to different hydrodynamic
equations (also see Refs. \citep{Kovtun:2012rj,Kovtun:2019hdm} for
the general discussion for these kinds of replacement in relativistic
hydrodynamics). Different with our Eqs. (\ref{eq:Six1stSolM1Lk-1}
- \ref{eq:Six1stSolM1Lk-7}), the dispersion relations computed with
the $q^{\mu}$ in Eq. (\ref{eq:RewriteSixQ}) are stable and satisfy
causality condition (\ref{eq:casuality_01}) in the rest frame under
certain conditions. However, they do not obey the causality condition
(\ref{eq:causality_02}) and the whole theory become acausal, e.g.,
one mode in Refs. \citep{Hattori:2019lfp,Sarwar:2022yzs,Daher:2022wzf}
is
\begin{equation}
\omega=i(\gamma^{\prime}+\gamma_{\perp})k^{2}\text{ as }k\rightarrow\infty,\label{eq:acasusal_01}
\end{equation}
breaks the causality condition (\ref{eq:causality_02}).

We now conclude that the first order spin hydrodynamics at static
equilibrium state are unstable and acausal in the rest frame. We do
not need to discuss the stability and causality of the first order
spin hydrodynamics in moving frames again.

\section{Minimal causal spin hydrodynamics \label{subsec:Minimal-causal-theory}}

In the previous section, we have shown that the first order spin hydrodynamics
in Landau frame are acausal and unstable. The acausal and unstable
theory is not physical, we therefore need to consider the second order
spin hydrodynamics in gradient expansion. In this section we follow
the idea of minimal causal extension in conventional hydrodynamics
and implement it to the spin hydrodynamics.

Up to now, there are two ways to establish causal hydrodynamics. The
first way is to add the second order corrections to the dissipative
terms, such as the M\"uller-Israel-Stewart (MIS) theory \citep{Israel:1979,Israel:1979wp}
or other related second order hydrodynamics. The MIS theory is a famous
causal conventional hydrodynamic theory up to $O(\partial^{2})$ in
gradient expansion. Here, we consider a relativistic dissipative hydrodynamics
with the bulk viscous pressure $\Pi$ only as an example to explain
why the MIS theory can be casual. The entropy current in MIS theory
is assumed to be \citep{Israel:1979wp,Muronga:2003ta,rezzolla2013relativistic}
\begin{equation}
\mathcal{S}^{\mu}=su^{\mu}-\frac{\mu}{T}\nu^{\mu}+\frac{1}{T}h^{\mu}-\frac{1}{2T}\beta_{0}u^{\mu}\Pi^{2}+...,
\end{equation}
where the coefficient $\beta_{0}>0$ and the ellipsis stands for other
possible $O(\partial^{2})$ terms. Then the second law of thermodynamics
$\partial_{\mu}\mathcal{S}^{\mu}\geq0$ leads to,
\begin{equation}
\tau_{\Pi}\frac{d}{d\tau}\Pi+\Pi=-\zeta\partial_{\mu}u^{\mu}+...,\label{eq:RelaxationTypeBulk}
\end{equation}
where $d/d\tau\equiv u^{\mu}\partial_{\mu}$, and $\tau_{\Pi}=\zeta\beta_{0}>0$
is defined as the relaxation time for the bulk viscous pressure. If
the $\tau_{\Pi}\rightarrow0$, the hydrodynamic equations reduce to
parabolic equations and become acausal. With a finite $\tau_{\Pi}$,
the hydrodynamic equations are hyperbolic and can be causal under
certain conditions \citep{Hiscock:1983zz,Olson:1990rzl,Denicol:2008ha,Pu:2009fj,Brito:2020nou}.
In linear modes analysis, the dispersion relations from Eq. (\ref{eq:RelaxationTypeBulk})
satisfy causal conditions (\ref{eq:casuality_01}, \ref{eq:causality_02})
when the relaxation time $\tau_{\Pi}$ is sufficiently large. The
second order constitutive equations for shear viscous tensor $\pi^{\mu\nu}$,
heat flow $h^{\mu}$ and heat current $\nu^{\mu}$ can be obtained
in a similar way. These equations represent evolution equations that
incorporate the respective relaxation time \citep{Israel:1979wp,Muronga:2003ta,rezzolla2013relativistic}.
Apart from the MIS theory, many other second order causal conventional
hydrodynamic theories, e.g., Baier-Romatschke-Son-Starinets-Stephanov
(BRSSS) theory \citep{Baier:2007ix} and the Denicol-Niemi-Molnar-Rischke
(DNMR) theory \citep{Denicol:2012cn}, have been established. All
of them contain the terms proportional to the relaxation times and
can be causal and stable under certain conditions \citep{Baier:2007ix,Floerchinger:2017cii,Bemfica:2020xym}.
Following these discussion, we can say that the key to recover the
causality of the theory is to introduce the terms proportional to
relaxation time.

Different with the above second order theories, the Bemfica-Disconzi-Noronha-Kovtun
(BDNK) \citep{Bemfica:2017wps,Bemfica:2019knx,Kovtun:2019hdm,Hoult:2020eho,Bemfica:2020zjp,Hoult:2021gnb}
is a first order hydrodynamic theory in general (fluid) frames. It
roughly says that one can choose some preferred frames to satisfy
the causality and stability conditions. Unfortunately, the commonly
used Landau or Eckart frame are not the preferred fluid frames in
the BDNK theory. Therefore, we will not discuss the spin hydrodynamics
in the BDNK theory in this work. We also notice that recent studies
in Ref. \citep{Weickgenannt:2023btk} discuss the casual spin hydrodynamics
in the first order similar to BDNK theory.\textcolor{blue}{{} }

In this work, we follow the basic idea in MIS, BRSSS, and DNMR theories
to construct a simplified causal spin hydrodynamics. Instead of considering
the complete second order spin hydrodynamics, we only analyze the
called ``minimal'' extended second order spin hydrodynamics. Here,
the word ``minimal'' means that we concentrate on the essential
terms in the second order of gradient expansion to get a causal theory
and neglect the other terms which do not contribute to the dispersion
relations in the linear modes analysis. As mentioned below Eq. (\ref{eq:RelaxationTypeBulk}),
the key to get the causal theory is to add the terms proportional
to the relaxation times similar to $\tau_{\Pi}d\Pi/d\tau$, in the
left hand side of Eq. (\ref{eq:RelaxationTypeBulk}). Following this
idea, the constitutive equations (\ref{eq:SixHeatCurrent}-\ref{eq:SixPhi})
in the minimal extended causal spin hydrodynamics can be rewritten
as,
\begin{eqnarray}
\tau_{q}\Delta^{\mu\nu}\frac{d}{d\tau}q_{\nu}+q^{\mu} & = & \lambda[T^{-1}\Delta^{\mu\alpha}\partial_{\alpha}T+(u\cdot\partial)u^{\mu}-4\omega^{\mu\nu}u_{\nu}],\label{eq:Type1q}\\
\tau_{\phi}\Delta^{\mu\alpha}\Delta^{\nu\beta}\frac{d}{d\tau}\phi_{\alpha\beta}+\phi^{\mu\nu} & = & 2\gamma_{s}\Delta^{\mu\alpha}\Delta^{\nu\beta}(\partial_{[\alpha}u_{\beta]}+2\omega_{\alpha\beta}),\label{eq:Type1phi}\\
\tau_{\pi}\Delta^{\alpha<\mu}\Delta^{\nu>\beta}\frac{d}{d\tau}\pi_{\alpha\beta}+\pi^{\mu\nu} & = & 2\eta\partial^{<\mu}u^{\nu>},\label{eq:Type1pi}\\
\tau_{\Pi}\frac{d}{d\tau}\Pi+\Pi & = & -\zeta\partial_{\mu}u^{\mu},\label{eq:Type1Pi}
\end{eqnarray}
where $\tau_{q},\tau_{\phi},\tau_{\pi}$ and $\tau_{\Pi}$ are positive
relaxation times for $q^{\mu},\phi^{\mu\nu},\pi^{\mu\nu},\Pi$, respectively.
Eqs. (\ref{eq:Type1pi},\ref{eq:Type1Pi}) are the same as those in
the conventional hydrodynamics\footnote{Another kinds of the minimal causal theory is discussed in Ref.\citep{Koide:2006ef,Denicol:2009yyt},
in which the extended dissipative terms can not be determined from
the entropy principle $\partial_{\mu}\mathcal{S}^{\mu}\geq0$.} \citep{Denicol:2008ha,Pu:2009fj,Brito:2020nou}. Recently, the second
order spin hydrodynamics similar to MIS theory has been introduced
in Ref. \citep{Biswas:2023qsw} by using the entropy principle. Our
minimal causal spin hydrodynamics can be regarded as a simplified
version of it. We also notice that in the Refs. \citep{Liu:2020ymh},
the authors have proposed the same expressions for $q^{\mu}$ and
$\phi^{\mu\nu}$ as presented in Eqs. (\ref{eq:Type1q}, \ref{eq:Type1phi})
for minimal causal spin hydrodynamics. 

Let us give some physical interpretation for Eqs. (\ref{eq:Type1q}-\ref{eq:Type1Pi}).
The nonzero relaxation times imply that the system requires time to
transition from a non-equilibrium state to an equilibrium state. 
In other words, the dissipative fluxes $\Pi$, $\pi^{\mu\nu}$, $q^{\mu}$,
and $\phi^{\mu\nu}$ do not undergo sudden transitions from nonzero
to zero \citep{Israel:1979wp,Muronga:2003ta}. As an example, we consider
the general solution for $\Pi$ \citep{Koide:2006ef} 
\begin{equation}
\Pi=\Pi_{0}e^{-(\tau-\tau_{0})/\tau_{\Pi}}-\int_{\tau_{0}}^{\tau}d\tau^{\prime}G(\tau,\tau^{\prime})\zeta\partial_{\mu}u^{\mu},\label{eq:Pi}
\end{equation}
where $\Pi_{0}$ is constant, and the Green's function is defined
as follows: $G(\tau,\tau^{\prime})=0$ for $\tau<\tau^{\prime}$,
$G(\tau,\tau^{\prime})=1/(2\tau_{\Pi})$ for $\tau=\tau^{\prime}$,
and $G(\tau,\tau^{\prime})=\frac{1}{\tau_{\Pi}}e^{-(\tau-\tau^{\prime})/\tau_{\Pi}}$
for $\tau>\tau^{\prime}$. The general solutions for $\pi^{\mu\nu}$,
$q^{\mu}$, and $\phi^{\mu\nu}$ in Eqs. (\ref{eq:Type1q}-\ref{eq:Type1pi})
share a structure similar to that  of Eq. (\ref{eq:Pi}). Now, we
assume that $\zeta\partial_{\mu}u^{\mu}$ jumps from nonzero to zero
at time $\tau_{0}$. Due to the nonzero relaxation time $\tau_{\Pi}$,
the solution (\ref{eq:Pi}) indicates that $\Pi$ cannot instantaneously
switch from a nonzero (non-equilibrium) value to zero (equilibrium).
However, if $\tau_{\Pi}=0$, the solution (\ref{eq:Pi}) reduces to
$\Pi=-\zeta\partial_{\mu}u^{\mu}$, and then $\Pi$ undergoes sudden
change from nonzero to zero and it thus causes acausality. Therefore,
to obtain a physical theory, we introduce the nonzero relaxation times
and treat $\Pi$, $\pi^{\mu\nu}$, $q^{\mu}$, and $\phi^{\mu\nu}$
as dynamical variables in Eqs. (\ref{eq:Type1q}-\ref{eq:Type1Pi}).
In principle, we can also consider the nonzero $\Sigma_{(1)}^{\lambda\mu\nu}$
in Eq. (\ref{eq:spindof6}), which might involve corrections similar
to the relaxation terms for $q^{\mu}$ and $\phi^{\mu\nu}$. In this
work, we concentrate on the simplest extension of the second-order
terms and leave the more general discussion for future research.

%2023.06.14

\section{Causality and stability analysis for minimal causal spin hydrodynamics
\label{sec:LinearMinimalCausalSpinHydro}}

In this section we analyze the causality and stability of the minimal
causal spin hydrodynamics. We use the similar notations in Sec. \ref{sec:CausalityStabilityFirstOrder},
i.e., for a physical quantity $X$, we use $X_{(0)}$ and $\delta X$
to denote the $X$ at the global equilibrium state and the small perturbations
of the quantity $X$, respectively. We adopt the independent perturbations
as 
\begin{equation}
\delta e,\ \delta u^{i},\ \delta S^{\mu\nu},\ \delta\Pi,\ \delta\pi^{ij},
\end{equation}
where $\delta\pi_{\ i}^{i}=0$ and $\delta\pi^{ij}=\delta\pi^{ji}$.

We first start from the spin hydrodynamics in the rest frame, i.e.,
$u_{(0)}^{\mu}=(1,0)$. The conservation equations $\partial_{\mu}\delta\Theta^{\mu\nu}=0$
and $\partial_{\lambda}\delta J^{\lambda\mu\nu}=0$ with the constitutive
equations (\ref{eq:Type1q} - \ref{eq:Type1Pi}) read, 
\begin{eqnarray}
0 & = & (\lambda^{\prime}c_{s}^{2}\partial^{i}+8\lambda\chi_{e}^{0i})\delta e+\lambda^{\prime}\partial_{0}\delta\vartheta^{i}+(2D_{b}-\tau_{q}\partial_{0}\partial_{0}-\partial_{0})\delta S^{0i},\label{eq:SpinDof6EQq}\\
0 & = & 8\gamma_{s}\chi_{e}^{ij}\delta e+2\gamma^{\prime}(\partial^{i}\delta\vartheta^{j}-\partial^{j}\delta\vartheta^{i})+(\tau_{\phi}\partial_{0}\partial_{0}+\partial_{0}+2D_{s})\delta S^{ij},\label{eq:SpinDof6EQphi}\\
0 & = & \tau_{\pi}\partial_{0}\delta\pi^{ij}+\delta\pi^{ij}-\gamma_{\perp}(\partial^{i}\delta\vartheta^{j}+\partial^{j}\delta\vartheta^{i}-\frac{2}{3}g^{ij}\partial_{k}\delta\vartheta^{k}),\label{eq:SpinDof6EQpi}\\
0 & = & \tau_{\Pi}\partial_{0}\delta\Pi+\delta\Pi+(\gamma_{\|}-\frac{4}{3}\gamma_{\perp})\partial_{i}\delta\vartheta^{i},\label{eq:SpinDof6EQPi}\\
0 & = & \partial_{0}\delta e+\partial_{i}\delta\vartheta^{i}+\frac{1}{2}\partial_{0}\partial_{i}\delta S^{0i},\label{eq:SpinDof6EQT1}\\
0 & = & -c_{s}^{2}\partial^{j}\delta e+\partial_{0}\delta\vartheta^{j}-\partial^{j}\delta\Pi+\partial_{i}\delta\pi^{ij}-\frac{1}{2}\partial_{0}\partial_{0}\delta S^{0j}-\frac{1}{2}\partial_{0}\partial_{i}\delta S^{ij},\label{eq:SpinDof6EQT2}
\end{eqnarray}
where $\chi_{b},\chi_{e}^{\mu\nu},\chi_{s},D_{s},D_{b},\delta\vartheta^{i},\lambda^{\prime},\gamma^{\prime},\gamma_{\perp},\gamma_{\|}$
are defined in Eqs. (\ref{eq:Notations0},\ref{eq:Notations1}) and
we have used the spin evolution equation (\ref{eq:SpinEvo}) to replace
$\delta q^{i}$ and $\delta\phi^{ij}$ by $\delta S^{\mu\nu}$, 
\begin{equation}
\delta q^{i}=\frac{1}{2}\partial_{0}\delta S^{0i},\quad\delta\phi^{ij}=-\frac{1}{2}\partial_{0}\delta S^{ij}.
\end{equation}

\subsection{Zero modes for the spin hydrodynamics with zero viscous effects \label{subsec:ZeroModes}}

Following the conventional hydrodynamics, we consider a fluid with
the dissipative terms $q^{\mu}$ and $\phi^{\mu\nu}$ only for simplicity,
i.e., we remove Eqs. (\ref{eq:SpinDof6EQpi}, \ref{eq:SpinDof6EQPi})
and take $\delta\Pi=0$ and $\delta\pi^{ij}=0$ in Eqs. (\ref{eq:SpinDof6EQq},
\ref{eq:SpinDof6EQphi}, \ref{eq:SpinDof6EQT1}, \ref{eq:SpinDof6EQT2}).
The detail of the calculation is shown in Appendix \ref{subsec:Analysis-for-I}.
The causality condition requires 
\begin{equation}
0\leq\frac{c_{s}^{2}(3\lambda^{\prime}+2\tau_{q})}{2\tau_{q}-\lambda^{\prime}}\leq1,\ 0\leq\frac{2\gamma^{\prime}\tau_{q}}{(2\tau_{q}-\lambda^{\prime})\tau_{\phi}}\leq1.\label{eq:casuality-zero-modes-1}
\end{equation}
The stability conditions give, 
\begin{equation}
\tau_{q}>\lambda^{\prime}/2,\ D_{s}>0,\ D_{b}<0,\ \chi_{e}^{0x}=0.\label{eq:stability-Q-2}
\end{equation}
The above conditions are derived from the small $k$ and large $k$
limits only. We can implement the Routh-Hurwitz criterion \citep{Gopal2006ControlSP,Gradshteuin:2007book,Kovtun:2019hdm,Bemfica:2019knx,Hoult:2020eho,Bemfica:2020zjp}
to prove that the condition (\ref{eq:stability-Q-2}) is sufficient
and necessary for stability. More discussion can be found in Appendix
\ref{sec:RHwithoutCouple}. 

Interestingly, there exist zero modes, i.e., $\omega=0$ for all $k$,
coming from Eq. (\ref{eq:SpinDof6EQT2}) with vanishing $\delta\Pi,\delta\pi^{ij}$.
Generally, the zero modes in the linear mode analysis do not mean
the perturbations are not decaying with time. It indicates that the
nonlinear modes should be included in Eq. (\ref{eq:SpinDof6EQT2})
if $\delta\Pi=\delta\pi^{ij}=0$. To continue our linear mode analysis,
we need to set non-vanishing $\delta\Pi,\delta\pi^{ij}$. 

\subsection{Causality analysis in the rest frame\label{subsec:Causality-analysis-in}}

Next, we substitute the plane wave solutions Eq. (\ref{eq:PlaneWave1})
and 
\begin{equation}
\delta\Pi=\delta\tilde{\Pi}e^{i\omega t-ikx},\ \delta\pi^{ij}=\delta\tilde{\pi}^{ij}e^{i\omega t-ikx},\label{eq:PlaneWave2}
\end{equation}
with $\delta\tilde{\Pi},\delta\tilde{\pi}^{ij}$, being constants,
into Eqs. (\ref{eq:SpinDof6EQq}-\ref{eq:SpinDof6EQT2}), and obtain
the matrix equation
\begin{eqnarray}
\mathcal{M}_{2}\delta\tilde{X}_{2} & = & 0,
\end{eqnarray}
where\textcolor{orange}{{} }$\delta\tilde{X}_{2}$ and $\mathcal{M}_{2}$
are given by 
\begin{eqnarray}
\delta\tilde{X}_{2} & \equiv & (\delta\tilde{e},\delta\tilde{\vartheta}^{x},\delta\tilde{S}^{0x},\delta\tilde{\Pi},\delta\tilde{\pi}^{xx},\delta\tilde{\vartheta}^{y},\delta\tilde{S}^{0y},\delta\tilde{S}^{xy},\delta\tilde{\pi}^{xy}\nonumber \\
 &  & ,\delta\tilde{\vartheta}^{z},\delta\tilde{S}^{0z},\delta\tilde{S}^{xz},\delta\tilde{\pi}^{xz},\delta\tilde{S}^{yz},\delta\tilde{\pi}^{yy},\delta\tilde{\pi}^{yz})^{\mathrm{T}}.
\end{eqnarray}
and 
\begin{eqnarray}
\mathcal{M}_{2} & = & \left(\begin{array}{cccc}
M_{4} & 0 & 0 & 0\\
A_{4} & M_{5} & 0 & 0\\
A_{5} & 0 & M_{5} & 0\\
A_{6} & 0 & 0 & M_{6}
\end{array}\right),\label{eq:Spindof6M2}
\end{eqnarray}
with 
\begin{eqnarray}
M_{4} & = & \left(\begin{array}{ccccc}
i\omega & -ik & \frac{1}{2}\omega k & 0 & 0\\
-ikc_{s}^{2} & i\omega & \frac{1}{2}\omega^{2} & -ik & -ik\\
ik\lambda^{\prime}c_{s}^{2}+8\lambda\chi_{e}^{0x} & i\omega\lambda^{\prime} & 2D_{b}+\tau_{q}\omega^{2}-i\omega & 0 & 0\\
0 & -ik(\gamma_{\|}-\frac{4}{3}\gamma_{\perp}) & 0 & i\omega\tau_{\Pi}+1 & 0\\
0 & -\frac{4}{3}ik\gamma_{\perp} & 0 & 0 & i\omega\tau_{\pi}+1
\end{array}\right),\\
M_{5} & = & \left(\begin{array}{cccc}
2ik\gamma^{\prime} & 0 & -\tau_{\phi}\omega^{2}+i\omega+2D_{s} & 0\\
i\omega & \frac{1}{2}\omega^{2} & -\frac{1}{2}\omega k & -ik\\
i\omega\lambda^{\prime} & 2D_{b}+\tau_{q}\omega^{2}-i\omega & 0 & 0\\
-ik\gamma_{\perp} & 0 & 0 & i\omega\tau_{\pi}+1
\end{array}\right),\label{eq:M5}\\
M_{6} & = & \left(\begin{array}{ccc}
-\tau_{\phi}\omega^{2}+i\omega+2D_{s} & 0 & 0\\
0 & i\omega\tau_{\pi}+1 & 0\\
0 & 0 & i\omega\tau_{\pi}+1
\end{array}\right).\label{eq:M6}
\end{eqnarray}
The submatrices $A_{4,5,6}$ in Eq. (\ref{eq:Spindof6M2}) are shown
in Appendix \ref{sec:Off-diagonal-submatrices-in}. If there exist
nonzero plane wave solutions, we have 
\begin{equation}
0=\det\mathcal{M}_{2}=\det M_{4}\cdot(\det M_{5})^{2}\cdot\det M_{6}.\label{eq:spindof6Minimal}
\end{equation}
We observe the zero modes in Eq. (\ref{eq:SpinDof6EQT2}) disappear.
It indicates that the current analysis is consistent with the assumption
of linear response. The dispersion relations $\omega=\omega(k)$ are
the solutions to the polynomial equation (\ref{eq:spindof6Minimal}). 

The $\det M_{6}=0$ gives, 
\begin{eqnarray}
\omega & = & \frac{i}{\tau_{\pi}},\label{eq:M6_1}\\
\omega & = & \frac{1}{2\tau_{\phi}}(i\pm\sqrt{8D_{s}\tau_{\phi}-1}),\label{eq:M6_2}
\end{eqnarray}
which are non-propagating modes or non-hydrodynamic modes.

In $k\rightarrow0$ limit, the $\det M_{4}=0$ and $\det M_{5}=0$
give  
\begin{eqnarray}
\omega & = & \frac{i}{\tau_{\pi}}+O(k),\label{eq:Six2Smallk-1}\\
\omega & = & \frac{i}{\tau_{\Pi}}+O(k),\label{eq:Six2Smallk-2}\\
\omega & = & \pm c_{s}k+\frac{i}{2}(\gamma_{\|}\mp4c_{s}\lambda\chi_{e}^{0x}D_{b}^{-1})k^{2}+O(k^{3}),\label{eq:Six2Smallk-3}\\
\omega & = & \left[i\pm\sqrt{-4D_{b}(2\tau_{q}-\lambda^{\prime})-1}\right](2\tau_{q}-\lambda^{\prime})^{-1}+O(k),\label{eq:Six2Smallk-4}\\
\omega & = & i\gamma_{\perp}k^{2}+O(k^{3}),\label{eq:Six2Smallk-5}\\
\omega & = & \frac{1}{2\tau_{\phi}}(i\pm\sqrt{8D_{s}\tau_{\phi}-1})+O(k),\label{eq:Six2Smallk-6}
\end{eqnarray}
where Eq. (\ref{eq:Six2Smallk-1}) and Eq. (\ref{eq:Six2Smallk-4})
are doubly degenerate. In large $k$ limit, we have
\begin{eqnarray}
\omega & = & -4iD_{b}\gamma_{\|}^{-1}\lambda^{\prime-1}k^{-2}+O(k^{-3}),\label{eq:Six2Largek-1}\\
\omega & = & \frac{3i\gamma_{\|}}{\tau_{\pi}(3\gamma_{\|}-4\gamma_{\perp})+4\gamma_{\perp}\tau_{\Pi}}+O(k^{-1}),\label{eq:Six2Largek-2}\\
\omega & = & c_{1}k+i\frac{c_{2}}{c_{3}}+O(k^{-1}),\label{eq:Six2Largek-3}\\
\omega & = & \pm\sqrt{\frac{2\tau_{q}(\gamma^{\prime}\tau_{\pi}+\gamma_{\perp}\tau_{\phi})}{(2\tau_{q}-\lambda^{\prime})\tau_{\pi}\tau_{\phi}}}k+ic_{4}+O(k^{-1}),\label{eq:Six2Largek-4}\\
\omega & = & \frac{i\pm\sqrt{-1-8D_{b}\tau_{q}}}{2\tau_{q}}+O(k^{-1}),\label{eq:Six2Largek-5}\\
\omega & = & \frac{i(\gamma^{\prime}+\gamma_{\perp})\pm c_{5}}{2(\gamma^{\prime}\tau_{\pi}+\gamma_{\perp}\tau_{\phi})}+O(k^{-1}),\label{eq:Six2Largek-6}
\end{eqnarray}
where the expressions of these $k$-independent coefficients $c_{1,2,3,4,5}$
are shown in Appendix \ref{sec:Definitions_ci}. The $\det M_{4}=0$
gives Eqs. (\ref{eq:Six2Smallk-1}-\ref{eq:Six2Smallk-4}) and Eqs.
(\ref{eq:Six2Largek-1}-\ref{eq:Six2Largek-3}), while $\det M_{5}=0$
gives Eqs. (\ref{eq:Six2Smallk-1},\ref{eq:Six2Smallk-4}-\ref{eq:Six2Smallk-6})
and Eqs. (\ref{eq:Six2Largek-4}-\ref{eq:Six2Largek-6}).

Now, let us analyze the causality conditions. From Eqs. (\ref{eq:Six2Largek-1}-\ref{eq:Six2Largek-6}),
we find that all modes in minimal causal spin hydrodynamics correspond
to finite propagation speed since $|\omega/k|$ is bounded as $k\rightarrow+\infty$.
Imposing Eq. (\ref{eq:casuality_01}) on the propagating modes in
Eqs. (\ref{eq:Six2Largek-3}-\ref{eq:Six2Largek-4}), the causality
requires,
\begin{equation}
0\leq\frac{b_{1}^{1/2}\pm(b_{1}-b_{2})^{1/2}}{6(2\tau_{q}-\lambda^{\prime})\tau_{\pi}\tau_{\Pi}}\leq1\text{ and }0\leq\frac{2\tau_{q}(\gamma^{\prime}\tau_{\pi}+\gamma_{\perp}\tau_{\phi})}{(2\tau_{q}-\lambda^{\prime})\tau_{\pi}\tau_{\phi}}\leq1,\label{eq:CforMSix}
\end{equation}
where $b_{1,2}$ are defined in Appendix \ref{sec:Definitions_ci}.
The causality conditions imply that the relaxation times $\tau_{q},\tau_{\pi},\tau_{\Pi},\tau_{\phi}$
cannot be arbitrarily small, which is consistent with the discussion
in Sec. \ref{subsec:Minimal-causal-theory}. We also notice that the
Eq. (\ref{eq:CforMSix}) reduces to Eq. (\ref{eq:casuality_condtion_q_phi})
when we take a smooth limit $\tau_{\pi},\tau_{\Pi},\gamma_{\perp},\gamma_{\parallel}\rightarrow0$.

\subsection{Non-trivial stability conditions in rest frame\label{subsec:Non-trivial-stability-conditions}}

The requirement of stability is non-trivial. Inserting Eq. (\ref{eq:Stablity_condition_01})
into Eqs. (\ref{eq:M6_1}-\ref{eq:Six2Largek-6}) yields,
\begin{eqnarray}
\tau_{q} & > & \lambda^{\prime}/2,\label{eq:SforMSix_01}\\
D_{s} & > & 0,\quad D_{b}<-4c_{s}\lambda\gamma_{\|}^{-1}\big|\chi_{e}^{0x}\big|\leq0,\label{eq:SforMSix_02}\\
b_{1} & > & b_{2}>0,\quad\frac{c_{2}}{c_{3}}>0.\label{eq:SforMSix_03}
\end{eqnarray}

The stability condition $\lambda^{\prime}<0$ in Eq. (\ref{eq:FirstSixUnstable})
for the first order spin hydrodynamics becomes $\lambda^{\prime}<2\tau_{q}$
in Eq.  (\ref{eq:SforMSix_01}). When the relaxation time $\tau_{q}$
is sufficiently large, the inequality $\lambda^{\prime}<2\tau_{q}$
is satisfied, and then the previous unstable modes are removed. We
also notice that the conditions (\ref{eq:SforMSix_01}, \ref{eq:SforMSix_02})
agree with Eq.  (\ref{eq:Stability-Q}) except the $\chi_{e}^{0x}=0$.
The strong constraint $\chi_{e}^{0x}=0$ is released in this case.

The satisfaction of the stability condition (\ref{eq:SforMSix_02})
relies on the specific equation of state governing $S^{\mu\nu}$ and
$\omega^{\mu\nu}$. In Ref.\citep{Daher:2022wzf}, it was found that
the stability condition (\ref{eq:SforMSix_02}) cannot be satisfied
if $\delta S^{\mu\nu}\sim T^{2}\delta\omega^{\mu\nu}$ \citep{Wang:2021ngp,Wang:2021wqq}.
In more general cases, we can have 
\begin{eqnarray}
u_{\mu}\delta\omega^{\mu\nu} & = & \chi_{1}u_{\mu}\delta S^{\mu\nu},\\
\Delta^{\mu\alpha}\Delta^{\nu\beta}\delta\omega_{\alpha\beta} & = & (\chi_{1}+\chi_{2})\Delta^{\mu\alpha}\Delta^{\nu\beta}\delta S_{\alpha\beta},
\end{eqnarray}
where $\chi_{1,2}$ are susceptibility corresponding to the $S^{0i}$
and $S^{ij}$ in the rest frame. In this case, according to the definitions
in Eqs. (\ref{eq:Notations0},\ref{eq:Notations1}), the stability
condition (\ref{eq:SforMSix_02}) is satisfied if $\chi_{2}>-\chi_{1}>0$.
Details can be found in Appendix \ref{sec:Discussions-about-}. Note
that the parameters $\chi_{1}$ and $\chi_{2}$ strongly depend on
the equation of state for $S^{\mu\nu}$ and $\omega^{\mu\nu}$. To
determine the equation of state, we need the microscopic theories,
and we will leave it for the future studies.

%2023.06.14

% stability in middle k region

\begin{figure}[t]
\noindent \begin{centering}
\includegraphics[scale=0.35]{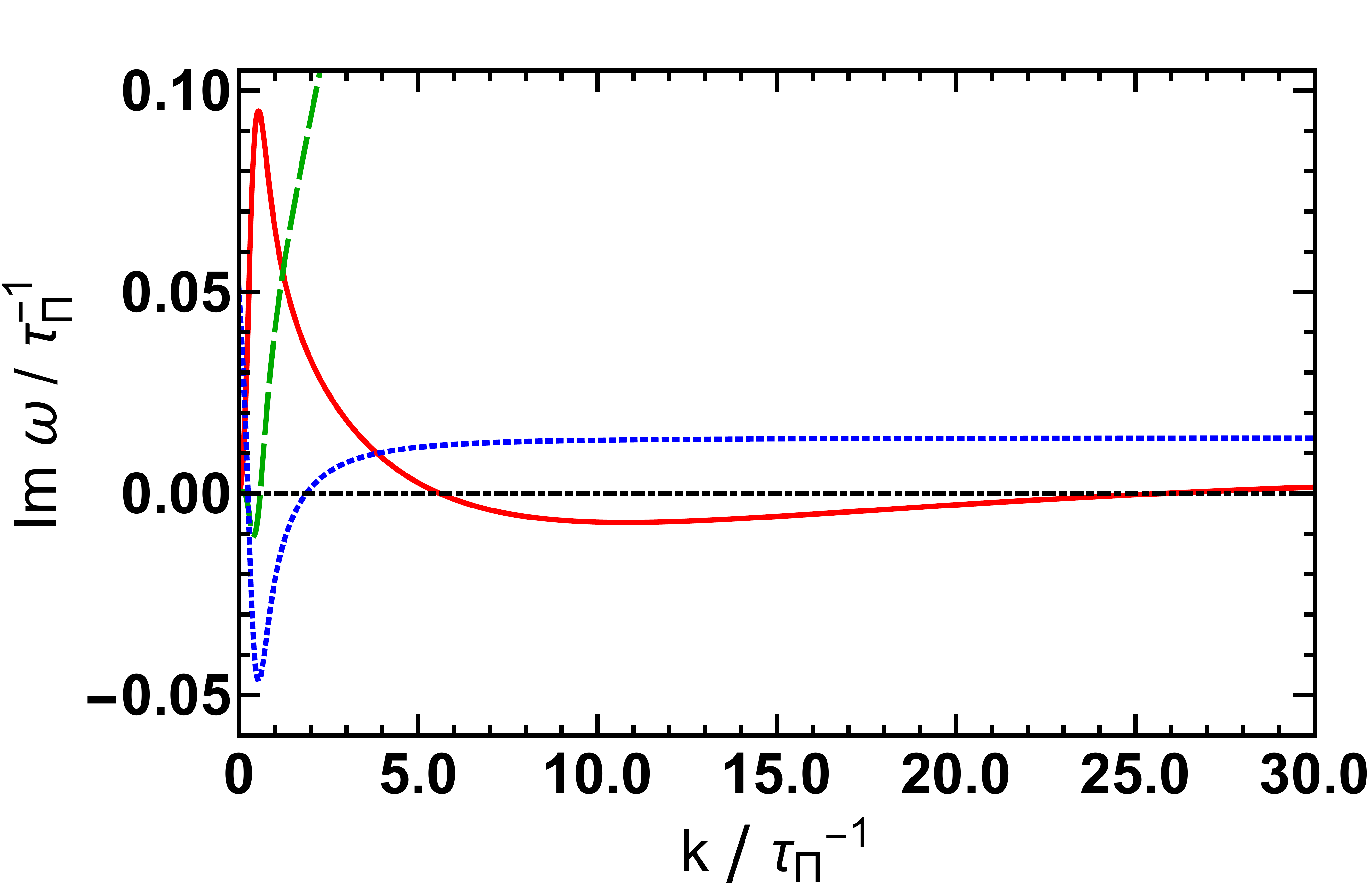}
\par\end{centering}
\caption{We plot the imaginary parts of $\omega\tau_{\Pi}$ as a function of
$k\tau_{\Pi}$ in three modes derived from $\det M_{4}=0$. The parameters
are chosen as in Eq. (\ref{eq:ParametersValue}), which satisfy the
causality and stability conditions Eqs. (\ref{eq:casuality_01}, \ref{eq:causality_02},
\ref{eq:Stablity_condition_01}). The soild, dashed and dotted lines
stand for three unstable modes. \label{fig:Imaginary-parts-of}}
\end{figure}

Another remarkable observation for the stability conditions is that
there exist unstable modes at finite $k$. Eqs. (\ref{eq:SforMSix_01},
\ref{eq:SforMSix_02}, \ref{eq:SforMSix_03}) are the stability conditions
in small $k$ and large $k$ limits only. We still need to study the
$\mathrm{Im}\ \omega$ in finite $k$ region. One analytic method,
named the Routh-Hurwitz criterion\textcolor{red}{{} }\citep{Gopal2006ControlSP,Gradshteuin:2007book,Kovtun:2019hdm,Bemfica:2019knx,Hoult:2020eho,Bemfica:2020zjp},
are usually implemented to study the sign of $\mathrm{Im}\ \omega$
in finite $k$ region. Unfortunately, $\det\mathcal{M}_{2}$ cannot
be reduced to the form that Routh-Hurwitz criterion applies, thus,
we analyze the behavior of $\mathrm{Im}\ \omega$ numerically instead
of the Routh-Hurwitz criterion. For a finite $k$, we find that $\mathrm{Im}\ \omega$
can be negative, even if all the conditions (\ref{eq:SforMSix_01},
\ref{eq:SforMSix_02}, \ref{eq:SforMSix_03}) are satisfied. In Fig.
\ref{fig:Imaginary-parts-of}, we present an example to show that
$\mathrm{Im}\ \omega$ can be negative for finite $k$. We choose
the parameters as,
\begin{eqnarray}
c_{s}=\frac{1}{\sqrt{3}}, & \  & \lambda\chi_{e}^{0x}=\frac{1}{8},\ \tau_{\pi}=4\tau_{\Pi},\ \tau_{\phi}=2\tau_{\Pi},\ \tau_{q}=10\tau_{\Pi},\ \lambda^{\prime}=\frac{1}{2}\tau_{\Pi},\nonumber \\
\gamma_{\|}=\frac{7}{10}\tau_{\Pi}, & \  & \gamma_{\perp}=\frac{1}{2}\tau_{\Pi},\ \gamma^{\prime}=\tau_{\Pi},\ D_{s}=\frac{1}{2\tau_{\Pi}},\ D_{b}=-\frac{1}{2\tau_{\Pi}}.\label{eq:ParametersValue}
\end{eqnarray}
It is straightforward to verify that the parameters in Eq. (\ref{eq:ParametersValue})
satisfy the stability and causality constraints (\ref{eq:casuality_01},
\ref{eq:causality_02}, \ref{eq:Stablity_condition_01}). We pick
up three modes derived from $\det M_{4}=0$. We observe that the $\textrm{Im }\omega$
at both small and large $k$ limits are positive, while it becomes
negative when $k\tau_{\Pi}\sim0.5$ and $k\tau_{\Pi}\sim10.0$, i.e.,
the modes are unstable in finite $k$ region.

We comment on the unstable modes at finite $k$. The unstable modes
in the minimal causal spin hydrodynamics are significantly different
with those in the conventional hydrodynamics. As discussed in Refs.
\citep{Denicol:2008ha,Pu:2009fj,Kovtun:2019hdm,Bemfica:2019knx,Brito:2020nou,Hoult:2020eho},
the stability conditions obtained in $k\rightarrow0$ and $k\rightarrow+\infty$
limits are sufficient to ensure the stability at any real $k$. However,
it looks failed in minimal causal spin hydrodynamics. It implies that
the conditions (\ref{eq:SforMSix_01}, \ref{eq:SforMSix_02}, \ref{eq:SforMSix_03})
are necessary but may not be sufficient. At last, it is still unclear
whether the unstable modes at finite $k$ indicate the fluid becomes
unstable or not.

\subsection{Causality and stability analysis for extended $q^{\mu}$ and $\phi^{\mu\nu}$
\label{subsec:Causality-and-stability-q-phi}}

In principle, we can introduce the coupling terms for $\Pi$, $\pi^{\mu\nu}$,
$q^{\mu}$, and $\phi^{\mu\nu}$ on the right-hand side of Eqs. (\ref{eq:Type1q}-\ref{eq:Type1phi}).
These terms will alter the linearized hydrodynamic equations and then
the causality and stability conditions can be changed. In the current
work, as the first step, we focus on the simplest coupling between
$q^{\mu}$ and $\phi^{\mu\nu}$, 
\begin{eqnarray}
\tau_{q}\Delta^{\mu\nu}\frac{d}{d\tau}q_{\nu}+q^{\mu} & = & \lambda\left(T^{-1}\Delta^{\mu\nu}\partial_{\nu}T+u^{\nu}\partial_{\nu}u^{\mu}-4u_{\nu}\omega^{\mu\nu}\right)+g_{1}\Delta^{\mu\nu}\partial^{\rho}\phi_{\nu\rho},\label{eq:Type1q-coupled}\\
\tau_{\phi}\Delta^{\mu\alpha}\Delta^{\nu\beta}\frac{d}{d\tau}\phi_{\alpha\beta}+\phi^{\mu\nu} & = & 2\gamma_{s}\Delta^{\mu\alpha}\Delta^{\nu\beta}\left(\partial_{[\alpha}u_{\beta]}+2\omega_{\alpha\beta}\right)+g_{2}\Delta^{\mu\alpha}\Delta^{\nu\beta}\partial_{[\alpha}q_{\beta]},\label{eq:Type1phi-coupled}
\end{eqnarray}
where $g_{1,2}$ are new transport coefficients describing the coupling
between $q^{\mu}$ and $\phi^{\mu\nu}$. For more general coupling
terms, one can refer to Ref. \citep{Biswas:2023qsw}. 

Following the same method, Eqs. (\ref{eq:SpinDof6EQq},\ref{eq:SpinDof6EQphi})
become, 
\begin{eqnarray}
0 & = & (\lambda^{\prime}c_{s}^{2}\partial^{i}+8\lambda\chi_{e}^{0i})\delta e+\lambda^{\prime}\partial_{0}\delta\vartheta^{i}+(2D_{b}-\tau_{q}\partial_{0}\partial_{0}-\partial_{0})\delta S^{0i}-g_{1}\partial_{j}\partial_{0}\delta S^{ij},\label{eq:SpinDof6EQq-coupled}\\
0 & = & 8\gamma_{s}\chi_{e}^{ij}\delta e+2\gamma^{\prime}(\partial^{i}\delta\vartheta^{j}-\partial^{j}\delta\vartheta^{i})+(\tau_{\phi}\partial_{0}\partial_{0}+\partial_{0}+2D_{s})\delta S^{ij}\nonumber \\
 &  & +\frac{1}{2}g_{2}\partial^{i}\partial_{0}\delta S^{0j}-\frac{1}{2}g_{2}\partial^{j}\partial_{0}\delta S^{0i}.\label{eq:SpinDof6EQphi-coupled}
\end{eqnarray}

We first consider the cases without viscous effects. The causality
condition (\ref{eq:casuality-zero-modes-1}) becomes 
\begin{equation}
0\leq\frac{c_{s}^{2}(3\lambda^{\prime}+2\tau_{q})}{2\tau_{q}-\lambda^{\prime}}\leq1,\ 0\leq\frac{m}{4(2\tau_{q}-\lambda^{\prime})\tau_{\phi}}\leq1,
\end{equation}
where $m$ is defined in Eq. (\ref{eq:def_m}). While the stability
condition (\ref{eq:stability-Q-2}) is changed to 
\begin{equation}
\tau_{q}>\lambda^{\prime}/2,\ D_{s}>0,\ D_{b}<0,\ \chi_{e}^{0x}=0,\quad m>8\gamma^{\prime}\left(\frac{2}{2\tau_{q}-\lambda^{\prime}}+\frac{1}{\tau_{\phi}}\right)^{-1}.\label{eq:stability_Q_3}
\end{equation}
Details can be found in Appendix \ref{subsec:Analysis-for-II}. We
implement the Routh-Hurwitz criterion \citep{Gopal2006ControlSP,Gradshteuin:2007book,Kovtun:2019hdm,Bemfica:2019knx,Hoult:2020eho,Bemfica:2020zjp}
again to prove that these conditions (\ref{eq:stability_Q_3}) are
sufficient and necessary for stability. Details for the proof can
be found in Appendix \ref{subsec:prove-case-II}. Similar to Sec.
\ref{subsec:ZeroModes}, we still find the zero modes coming from
Eq. (\ref{eq:SpinDof6EQT2}). 

Therefore, we need to consider the non-vanishing viscous effects.
Now, the sub-matrix $M_{5}$ shown in Eq. (\ref{eq:M5}) is replaced
with 
\begin{eqnarray}
M_{5} & = & \left(\begin{array}{cccc}
2ik\gamma^{\prime} & -\frac{1}{4}g_{2}\omega k & -\tau_{\phi}\omega^{2}+i\omega+2D_{s} & 0\\
i\omega & \frac{1}{2}\omega^{2} & -\frac{1}{2}\omega k & -ik\\
i\omega\lambda^{\prime} & 2D_{b}+\tau_{q}\omega^{2}-i\omega & g_{1}\omega k & 0\\
-ik\gamma_{\perp} & 0 & 0 & i\omega\tau_{\pi}+1
\end{array}\right),\label{eq:M5-coupled}
\end{eqnarray}
while other sub-matrices are unaffected by $g_{1}$ and $g_{2}$.
The Eqs. (\ref{eq:Six2Largek-4}-\ref{eq:Six2Largek-6}) become,
\begin{eqnarray}
\omega & = & \pm\sqrt{\frac{f+f^{\prime}}{8(2\tau_{q}-\lambda^{\prime})\tau_{\pi}\tau_{\phi}}}k+i\frac{f+f^{\prime}}{4(2\tau_{q}-\lambda^{\prime})\tau_{\pi}\tau_{\phi}}c_{6}+\mathcal{O}(k^{-1}),\label{eq:Six2Largek-4-Coupled}\\
\omega & = & \pm\sqrt{\frac{f-f^{\prime}}{8(2\tau_{q}-\lambda^{\prime})\tau_{\pi}\tau_{\phi}}}k+i\frac{f-f^{\prime}}{4(2\tau_{q}-\lambda^{\prime})\tau_{\pi}\tau_{\phi}}c_{7}+\mathcal{O}(k^{-1}),\label{eq:Six2Largek-5-Coupled}\\
\omega & = & \pm4\sqrt{\frac{-D_{b}D_{s}}{g_{1}g_{2}}}k^{-1}+4i\frac{[D_{s}\gamma_{\perp}-D_{b}(\gamma_{\perp}+\gamma^{\prime})]}{g_{1}g_{2}\gamma_{\perp}}k^{-2}+\mathcal{O}(k^{-3}),\label{eq:Six2Largek-6-Coupled}
\end{eqnarray}
where the definitions of $f,f^{\prime},c_{6}$, and $c_{7}$ are defined
in Appendix \ref{sec:Definitions_ci}.

From these new dispersion relations, we obtain causality conditions,
\begin{equation}
0\leq\frac{b_{1}^{1/2}\pm(b_{1}-b_{2})^{1/2}}{6(2\tau_{q}-\lambda^{\prime})\tau_{\pi}\tau_{\Pi}}\leq1\text{ and }0\leq\frac{f\pm f^{\prime}}{8(2\tau_{q}-\lambda^{\prime})\tau_{\pi}\tau_{\phi}}\leq1,\label{eq:CforMSix-coupled}
\end{equation}
which reproduce Eq. (\ref{eq:CforMSix}) when $g_{1},g_{2}\rightarrow0$.

\begin{figure}[t]
\noindent \begin{centering}
\includegraphics[scale=0.35]{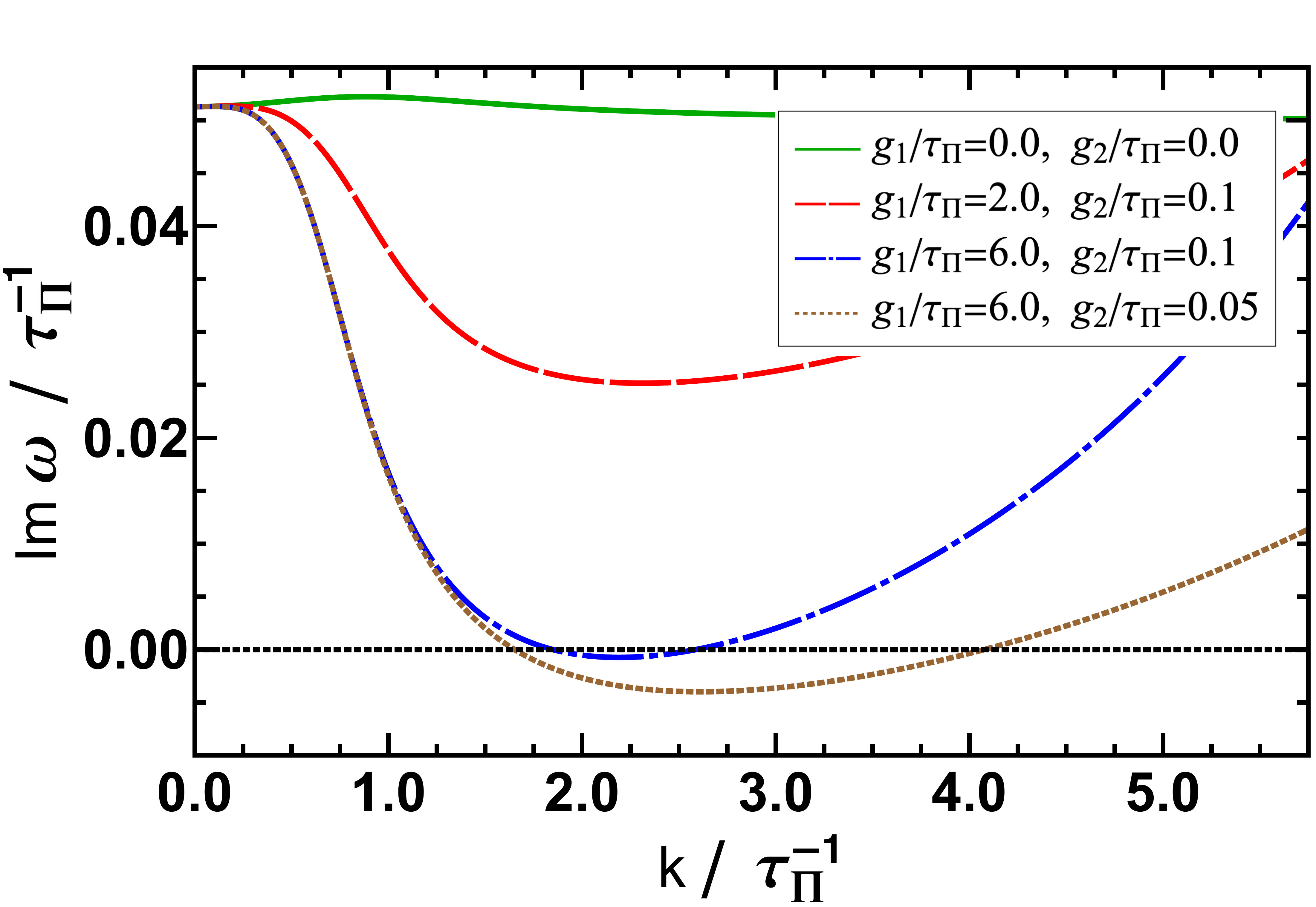}
\par\end{centering}
\caption{Imaginary parts of $\omega\tau_{\Pi}$ as a function of $k\tau_{\Pi}$
in one mode derived from $\det M_{5}=0$. The green solid, red dashed,
blue dash-dotted and brown dotted lines stand for the results with
$(g_{1}/\tau_{\Pi},g_{2}/\tau_{\Pi})=(0.0,0.0),(2.0,0.1),(6.0,0.1),(6.0,0.05)$.
Other parameters are also chosen as in Eq. (\ref{eq:ParametersValue}).
\label{fig:2}}
\end{figure}

Similarly, the stability conditions are given by 
\begin{eqnarray}
\tau_{q}-\frac{\lambda^{\prime}}{2} & > & 0,\label{eq:SforMSix_01-coupled}\\
D_{s}>0,\quad-4c_{s}\lambda\gamma_{\|}^{-1}\big|\chi_{e}^{0x}\big|-D_{b} & > & 0,\label{eq:SforMSix_02-coupled}\\
b_{1}>b_{2}>0,\quad\frac{c_{2}}{c_{3}} & > & 0,\label{eq:SforMSix_03-coupled}\\
g_{1}g_{2}>0,\quad f>0,\quad f^{\prime} & > & 0,\label{eq:SforMSix_04-coupled}\\
\textrm{Re}\ c_{6}>0,\quad\textrm{Re}\ c_{7} & > & 0.\label{eq:SforMSix_05-coupled}
\end{eqnarray}
Unfortunately, we find that the extended $q^{\mu}$ and $\phi^{\mu\nu}$
cannot remove the unstable modes at finite $k$ coming from $\textrm{det}M_{4}=0$.
We choose the parameters satisfying the causality conditions (\ref{eq:CforMSix-coupled})
and stability conditions (\ref{eq:SforMSix_01-coupled} - \ref{eq:SforMSix_05-coupled}),
and consider the influence on dispersion relations of $g_{1},g_{2}$.
For simplicity, we choose the parameters as the same as in Eq. (\ref{eq:ParametersValue})
with $(g_{1}/\tau_{\Pi},g_{2}/\tau_{\Pi})=(0.0,0.0),(2.0,0.1),(6.0,0.1),(6.0,0.05)$.
We find that one modes from $\det M_{5}=0$ becomes unstable at finite
$k$ with $(g_{1}/\tau_{\Pi},g_{2}/\tau_{\Pi})=(6.0,0.1),(6.0,0.05)$
as shown in Fig. \ref{fig:2}. 

As a brief summary, the extended $q^{\mu}$ and $\phi^{\mu\nu}$ can
modify the causality and stability conditions, but cannot remove the
zero modes when we turn off other dissipative effects. The unstable
modes at finite $k$ cannot be cured by the extended $q^{\mu}$ and
$\phi^{\mu\nu}$. 

\subsection{Causality and stability in moving frames}

Let us briefly discuss the causality and stability of the minimal
causal spin hydrodynamics in moving frames.

For the causality in a moving frame, we refer to the studies in Refs.
\citep{Kovtun:2019hdm,Hoult:2023clg,Wang:2023csj}. The authors in
Refs. \citep{Kovtun:2019hdm,Hoult:2023clg,Wang:2023csj} have studied
the dispersion relations at large $k$ limit in moving frames and
demonstrate that the system is causal in moving frames if it is causal
in the rest frame. Thus, the minimal causal spin hydrodynamics is
causal in moving frames when the causality condition (\ref{eq:CforMSix})
in the rest frame are satisfied.

For the stability, it has also been proved that if a causal theory
is unstable in the rest frame, then it is also unstable in moving
frames (see Theorem 2 of Ref. \citep{Gavassino:2021owo}). We now
apply this theorem to the minimal causal spin hydrodynamics. If the
equation of state gives $\delta\omega^{\mu\nu}=\chi_{1}\delta S^{\mu\nu}$
with constant $\chi_{1}$, the minimal causal spin hydrodynamics will
be unstable in moving frames since it has unstable modes in the rest
frame. For more general cases, the stability of the theory in both
moving frames and the rest frame depends on the equation of state
for $S^{\mu\nu}$ and $\omega^{\mu\nu}$.

In summary, the minimal causal spin hydrodynamics is causal in any
reference frame when Eq. (\ref{eq:CforMSix}) is fulfilled. Hence,
we have solved the problem of acausality by introducing the minimal
causal spin hydrodynamics. However, the stability of minimal causal
spin hydrodynamics remains unclear. Our findings indicate that the
validity of the stability condition (\ref{eq:SforMSix_02}) is highly
contingent upon the equation of state governing spin density and spin
chemical potential. Moreover, we also find that the stability conditions
(\ref{eq:SforMSix_01}, \ref{eq:SforMSix_02}, \ref{eq:SforMSix_03})
obtained at $k\rightarrow0$ and $k\rightarrow+\infty$ are necessary
but not sufficient.

\section{Conclusion \label{sec:Conclusion}}

In this work, we investigate the linear causality and stability of
the spin hydrodynamics proposed in Refs. \citep{Hattori:2019lfp,Fukushima:2020ucl}. 

In linear modes analysis, we consider perturbations to the spin hydrodynamics
near the static equilibrium. We obtain the dispersion relations $\omega=\omega(k)$
and analyze the all possible modes. The results show the stability
condition (\ref{eq:FirstSixUnstable}) cannot be fulfilled. Moreover,
the value of $|\omega/k|$ in Eqs. (\ref{eq:Six1stSolM1Lk-2}-\ref{eq:Six1stSolM1Lk-4})
is unbounded, which violates the causality condition (\ref{eq:causality_02}).
In Refs.\citep{Hattori:2019lfp,Sarwar:2022yzs,Daher:2022wzf}, the
expression of $q^{\mu}$ are modified by using the equation of motion
for the fluid. We emphasize that the first order spin hydrodynamics
in Refs. \citep{Hattori:2019lfp,Sarwar:2022yzs,Daher:2022wzf} are
still acausal since one mode shown in Eq. (\ref{eq:acasusal_01})
breaks the causality condition (\ref{eq:causality_02}). We conclude
that the spin hydrodynamics in the first order of gradient expansion
are acausal and unstable.

We then follow the basic idea in MIS, BRSSS, and DNMR theories and
consider the minimal causal spin hydrodynamics. The constitutive equations
(\ref{eq:SixHeatCurrent}-\ref{eq:SixPhi}) in a minimal extended
causal spin hydrodynamics are replaced by Eqs.  (\ref{eq:Type1q}-\ref{eq:Type1Pi}).
One can view it as a natural extension of the first order spin hydrodynamics
or a simplified version of the complete second order spin hydrodynamics
\citep{Biswas:2023qsw}.\textcolor{blue}{{} }We investigate the causality
and stability for this minimal causal spin hydrodynamics. We analyze
the causality and stability for dissipative fluids with $q^{\mu}$
and $\phi^{\mu\nu}$ only and find the zero modes in the linear modes
analysis. This suggests that linear mode analysis is inadequate in
this case. Therefore, we consider dissipative spin fluids with shear
viscous tensor and bulk viscous pressure. 

For causality, we find that the modes with infinite speed disappear
and all modes are causal in the rest frame if the conditions in Eq.
(\ref{eq:CforMSix}) are fulfilled. Following the statement in Refs.
\citep{Kovtun:2019hdm,Hoult:2023clg,Wang:2023csj}, we comment that
the minimal causal spin hydrodynamics are causal in any reference
frame when the conditions (\ref{eq:CforMSix}) are fulfilled. 

For the stability, although we obtain the stability conditions in
Eqs. (\ref{eq:SforMSix_01}, \ref{eq:SforMSix_02}, \ref{eq:SforMSix_03})
from the constraints in the $k\rightarrow0$ and $k\rightarrow+\infty$
limits, the stability of the theory in both moving frames and the
rest frame remains unclear. Two kinds of problems can lead to instabilities.
The first one is related to stability condition (\ref{eq:SforMSix_02}).
Interestingly, we prove that the coefficients $D_{s},D_{b}$ do not
obey the stability condition (\ref{eq:SforMSix_02}) if the equation
of state $S^{\mu\nu}\sim T^{2}\omega^{\mu\nu}$ is adopted. In more
general cases, the fulfillment of the stability condition (\ref{eq:SforMSix_02})
hinges on the specific equations of state. One has to assess the condition
(\ref{eq:SforMSix_02}) on a case-by-case basis. Surprisingly, different
with the conventional hydrodynamics, we find that the stability condition
(\ref{eq:Stablity_condition_01}) breaks at finite $k$ as shown in
Fig. \ref{fig:Imaginary-parts-of}. It implies that the conditions
(\ref{eq:SforMSix_01}, \ref{eq:SforMSix_02}, \ref{eq:SforMSix_03})
are necessary but may not be sufficient.

We also considered the extended $q^{\mu}$ and $\phi^{\mu\nu}$, in
which the $q^{\mu}$ and $\phi^{\mu\nu}$ are coupled in the second
order constitutive equations. The causality and stability conditions
are modified in this case. However, in dissipative fluids with $q^{\mu}$
and $\phi^{\mu\nu}$ only the zero modes cannot be removed. The unstable
modes at finite wavelength are still there. 

We conclude that the spin hydrodynamics in the first order of gradient
expansion, proposed in Refs. \citep{Hattori:2019lfp,Fukushima:2020ucl},
are always acausal and unstable. The minimal causal extension of it
makes the theory be causal in the sense of Eqs. (\ref{eq:casuality_01},\ref{eq:causality_02}).
However, the linear stability of the minimal causal spin hydrodynamics
remains unclear. The studies beyond the linear modes analysis may
provide us a better and clear answer to the problem of stability. 
\begin{acknowledgments}
We thank Francesco Becattini, Matteo Buzzegoli, Asaad Daher, Xu-Guang
Huang, Jin Hu, Masoud Shokri and David Wagner for helpful discussion
during the 7th International Conference on Chirality, Vorticity and
Magnetic Field in Heavy Ion Collisions. This work is supported in
part by the National Key Research and Development Program of China
under Contract No. 2022YFA1605500, by the Chinese Academy of Sciences
(CAS) under Grants No. YSBR-088 and by National Nature Science Foundation
of China (NSFC) under Grants No. 12075235 and No. 12135011.
\end{acknowledgments}

\appendix

\section{Off-diagonal submatrices in Eqs. (\ref{eq:Six1stFourierMatrix},
\ref{eq:Spindof6M2}) \label{sec:Off-diagonal-submatrices-in}}

In this appendix, we list all the off-diagonal submatrices introduced
in Eqs. (\ref{eq:Six1stFourierMatrix},\ref{eq:Spindof6M2}):

\begin{equation}
A_{1}\equiv\left(\begin{array}{ccc}
-4i(\omega\lambda\chi_{e}^{0y}+k\gamma_{s}\chi_{e}^{xy}) & 0 & 0\\
8\lambda\chi_{e}^{0y} & 0 & 0\\
8\gamma_{s}\chi_{e}^{xy} & 0 & 0
\end{array}\right),\ A_{2}\equiv\left(\begin{array}{ccc}
-4i(\omega\lambda\chi_{e}^{0z}+k\gamma_{s}\chi_{e}^{xz}) & 0 & 0\\
8\lambda\chi_{e}^{0z} & 0 & 0\\
8\gamma_{s}\chi_{e}^{xz} & 0 & 0
\end{array}\right),
\end{equation}
\begin{equation}
A_{3}=\left(\begin{array}{ccc}
8\gamma_{s}\chi_{e}^{yz}, & 0, & 0\end{array}\right),\ A_{4}=\left(\begin{array}{ccccc}
8\gamma_{s}\chi_{e}^{xy} & 0 & 0 & 0 & 0\\
0 & 0 & 0 & 0 & 0\\
8\lambda\chi_{e}^{0y} & 0 & 0 & 0 & 0\\
0 & 0 & 0 & 0 & 0
\end{array}\right),\ A_{5}=\left(\begin{array}{ccccc}
8\gamma_{s}\chi_{e}^{xz} & 0 & 0 & 0 & 0\\
0 & 0 & 0 & 0 & 0\\
8\lambda\chi_{e}^{0z} & 0 & 0 & 0 & 0\\
0 & 0 & 0 & 0 & 0
\end{array}\right),
\end{equation}
\begin{equation}
A_{6}=\left(\begin{array}{ccccc}
2\gamma_{s}\chi_{e}^{yz} & 0 & 0 & 0 & 0\\
0 & \frac{2}{3}ik\gamma_{\perp} & 0 & 0 & 0\\
0 & 0 & 0 & 0 & 0
\end{array}\right).
\end{equation}

\section{Definitions of coefficients \label{sec:Definitions_ci}}

The coefficients introduced in Eqs. (\ref{eq:Six2Largek-1}-\ref{eq:Six2Largek-6})
are defined as follows, 

\begin{eqnarray}
c_{1} & = & \sqrt{\frac{b_{1}^{1/2}\pm(b_{1}-b_{2})^{1/2}}{6(2\tau_{q}-\lambda^{\prime})\tau_{\pi}\tau_{\Pi}}},\textrm{or }-\sqrt{\frac{b_{1}^{1/2}\pm(b_{1}-b_{2})^{1/2}}{6(2\tau_{q}-\lambda^{\prime})\tau_{\pi}\tau_{\Pi}}},\\
c_{2} & = & -3c_{1}^{4}[2\tau_{\pi}\tau_{\Pi}+(2\tau_{q}-\lambda^{\prime})(\tau_{\pi}+\tau_{\Pi})]+48c_{1}^{3}\lambda\chi_{e}^{0x}\tau_{\pi}\tau_{\Pi}-3c_{s}^{2}\gamma_{\|}\lambda^{\prime}\nonumber \\
 &  & +c_{1}^{2}\{6\gamma_{\|}\tau_{q}+(6\gamma_{\|}-8\gamma_{\perp})\tau_{\pi}+8\gamma_{\perp}\tau_{\Pi}+3c_{s}^{2}[2\tau_{\pi}\tau_{\Pi}+(3\lambda^{\prime}+2\tau_{q})(\tau_{\pi}+\tau_{\Pi})]\}\nonumber \\
 &  & -8c_{1}\lambda\chi_{e}^{0x}[(3\gamma_{\|}-4\gamma_{\perp})\tau_{\pi}+4\gamma_{\perp}\tau_{\Pi}],\\
c_{3} & = & -2c_{s}^{2}\lambda^{\prime}[(3\gamma_{\|}-4\gamma_{\perp})\tau_{\pi}+4\gamma_{\perp}\tau_{\Pi}]-18c_{1}^{4}(2\tau_{q}-\lambda^{\prime})\tau_{\pi}\tau_{\Pi}\nonumber \\
 &  & +4c_{1}^{2}[3c_{s}^{2}(3\lambda^{\prime}+2\tau_{q})\tau_{\pi}\tau_{\Pi}+2(3\gamma_{\|}-4\gamma_{\perp})\tau_{q}\tau_{\pi}+8\gamma_{\perp}\tau_{q}\tau_{\Pi}],\\
c_{4} & = & \frac{\gamma_{\perp}[\tau_{q}(2\tau_{q}-\lambda^{\prime})+\lambda^{\prime}\tau_{\pi}]\tau_{\phi}^{2}+\gamma^{\prime}\tau_{\pi}^{2}[\tau_{q}(2\tau_{q}-\lambda^{\prime})+\lambda^{\prime}\tau_{\phi}]}{2(2\tau_{q}-\lambda^{\prime})\tau_{q}\tau_{\pi}\tau_{\phi}(\gamma^{\prime}\tau_{\pi}+\gamma_{\perp}\tau_{\phi})},\\
c_{5} & = & \sqrt{8D_{s}\gamma_{\perp}(\gamma^{\prime}\tau_{\pi}+\gamma_{\perp}\tau_{\phi})-(\gamma^{\prime}+\gamma_{\perp})^{2}},
\end{eqnarray}
where 
\begin{eqnarray}
b_{1} & = & \{8\gamma_{\perp}\tau_{q}\tau_{\Pi}+\tau_{\pi}[2\tau_{q}(3\gamma_{\|}-4\gamma_{\perp})+3\tau_{\Pi}c_{s}^{2}(3\lambda^{\prime}+2\tau_{q})]\}^{2},\\
b_{2} & = & 12c_{s}^{2}\lambda^{\prime}(2\tau_{q}-\lambda^{\prime})\tau_{\pi}\tau_{\Pi}[\tau_{\pi}(3\gamma_{\|}-4\gamma_{\perp})+4\gamma_{\perp}\tau_{\Pi}].
\end{eqnarray}
The coefficients used in Eqs. (\ref{eq:Six2Largek-4-Coupled}-\ref{eq:Six2Largek-6-Coupled})
are given by 
\begin{eqnarray}
f & = & m\tau_{\pi}+8\gamma_{\perp}\tau_{q}\tau_{\phi},\\
f^{\prime} & = & \{-32g_{1}g_{2}\gamma_{\perp}(2\tau_{q}-\lambda^{\prime})\tau_{\pi}\tau_{\phi}+(m\tau_{\pi}+8\gamma_{\perp}\tau_{q}\tau_{\phi})^{2}\}^{1/2},\\
c_{6} & = & -\frac{1}{(f^{\prime2}+fd^{1/2})}\left[m\tau_{\pi}(2\tau_{q}-\lambda^{\prime})(\tau_{\phi}-\tau_{\pi})\right.\nonumber \\
 &  & +8\gamma_{\perp}\tau_{q}\tau_{\phi}(\tau_{q}-\tau_{\pi})(\lambda^{\prime}-2\tau_{\phi})+16\gamma^{\prime}\tau_{\pi}^{2}\tau_{\phi}(2\tau_{q}-\lambda^{\prime})\nonumber \\
 &  & \left.-f^{\prime}(2\tau_{q}-\lambda^{\prime})(\tau_{\pi}+\tau_{\phi})+2\tau_{\pi}\tau_{\phi}(-m\tau_{\pi}-8\gamma_{\perp}\lambda^{\prime}\tau_{\phi}+8\gamma^{\prime}\tau_{q}^{2}-f^{\prime})\right],\\
c_{7} & = & -\frac{c_{72}}{c_{71}},
\end{eqnarray}
where 
\begin{eqnarray}
m & = & 2g_{1}g_{2}+8g_{1}\gamma^{\prime}+g_{2}\lambda^{\prime}+8\gamma^{\prime}\tau_{q},\label{eq:def_m}\\
c_{71} & = & -4g_{1}^{2}(g_{2}+4\gamma^{\prime})^{2}\tau_{\pi}^{2}+[g_{2}\lambda^{\prime}\tau_{\pi}+8\tau_{q}(\gamma^{\prime}\tau_{\pi}+\gamma_{\perp}\tau_{\phi})](-g_{2}\lambda^{\prime}\tau_{\pi}-8\gamma^{\prime}\tau_{q}\tau_{\pi}\nonumber \\
 &  & -8\gamma_{\perp}\tau_{q}\tau_{\phi}+d^{1/2})-2g_{1}\tau_{\pi}\{2g_{2}^{2}\lambda^{\prime}\tau_{\pi}+g_{2}[8\gamma^{\prime}(\lambda^{\prime}+2\tau_{q})\tau_{\pi}+16\gamma_{\perp}\lambda^{\prime}\tau_{\phi}\nonumber \\
 &  & -16\gamma_{\perp}\tau_{q}\tau_{\phi}-d^{1/2}]-4\gamma^{\prime}(16\gamma^{\prime}\tau_{q}\tau_{\pi}+16\gamma_{\perp}\tau_{q}\tau_{\phi}-d^{1/2})\},\\
c_{72} & = & -f^{\prime}(2\tau_{q}-\lambda^{\prime})(\tau_{\pi}+\tau_{\phi})+m\tau_{\pi}(2\tau_{q}-\lambda^{\prime})(\tau_{\pi}-\tau_{\phi})-16\gamma^{\prime}\tau_{\pi}^{2}\tau_{\phi}(2\tau_{q}-\lambda^{\prime})\nonumber \\
 &  & -8\gamma_{\perp}\tau_{q}\tau_{\phi}(2\tau_{q}-\lambda^{\prime})(\tau_{\pi}-\tau_{\phi})+\tau_{\pi}\tau_{\phi}(-2f^{\prime}+2m\tau_{\pi}-16\gamma_{\perp}\tau_{q}\tau_{\phi}+16\gamma_{\perp}\lambda^{\prime}\tau_{\phi}),\\
d & = & 4g_{1}^{2}(g_{2}+4\gamma^{\prime})^{2}\tau_{\pi}^{2}+[g_{2}\lambda^{\prime}\tau_{\pi}+8\tau_{q}(\gamma^{\prime}\tau_{\pi}+\gamma_{\perp}\tau_{\phi})]^{2}+4g_{1}\tau_{\pi}[g_{2}^{2}\lambda^{\prime}\tau_{\pi}\nonumber \\
 &  & \quad+4g_{2}\gamma^{\prime}(\lambda^{\prime}+2\tau_{q})\tau_{\pi}+8g_{2}\gamma_{\perp}(\lambda^{\prime}-\tau_{q})\tau_{\phi}+32\gamma^{\prime}\tau_{q}(\gamma^{\prime}\tau_{\pi}+\gamma_{\perp}\tau_{\phi})].
\end{eqnarray}

\section{Causality and stability of the minimal causal spin hydrodynamics
with $q^{\mu}$ and $\phi^{\mu\nu}$ only \label{sec:CausalityStabilityqphionly}}

In this appendix, we study the causality and stability of minimal
causal spin hydrodynamics, considering only $q^{\mu}$ and $\phi^{\mu\nu}$.
We will discuss two different cases. We name the system in which $q^{\mu}$
and $\phi^{\mu\nu}$ are not coupled, as depicted in Eqs. (\ref{eq:Type1q},\ref{eq:Type1phi})
as case I. Conversely, we name the system $q^{\mu}$ and $\phi^{\mu\nu}$
are coupled, as described by Eqs. (\ref{eq:Type1q-coupled},\ref{eq:Type1phi-coupled})
as case II. 

\subsection{Analysis for the case I \label{subsec:Analysis-for-I}}

We let $\delta\Pi=\delta\pi^{ij}=0$ in Eqs. (\ref{eq:SpinDof6EQq},
\ref{eq:SpinDof6EQphi}, \ref{eq:SpinDof6EQT1}, \ref{eq:SpinDof6EQT2})
and remove Eqs. (\ref{eq:SpinDof6EQpi}, \ref{eq:SpinDof6EQPi}).
Then we substitute the plane wave solutions Eq. (\ref{eq:PlaneWave1})
into Eqs. (\ref{eq:SpinDof6EQq}, \ref{eq:SpinDof6EQphi}, \ref{eq:SpinDof6EQT1},
\ref{eq:SpinDof6EQT2}) and derive 
\begin{eqnarray}
\mathcal{M}_{2}^{\prime}\delta\tilde{X}_{2}^{\prime} & = & 0,
\end{eqnarray}
where $\delta\tilde{X}_{2}^{\prime}$ and $\mathcal{M}_{2}^{\prime}$
are given by 
\begin{eqnarray}
\delta\tilde{X}_{2}^{\prime} & \equiv & (\delta\tilde{e},\delta\tilde{\vartheta}^{x},\delta\tilde{S}^{0x},\delta\tilde{\vartheta}^{y},\delta\tilde{S}^{0y},\delta\tilde{S}^{xy},\delta\tilde{\vartheta}^{z},\delta\tilde{S}^{0z},\delta\tilde{S}^{xz},\delta\tilde{S}^{yz})^{\mathrm{T}},
\end{eqnarray}
and 
\begin{eqnarray}
\mathcal{M}_{2}^{\prime} & \equiv & \left(\begin{array}{cccc}
M_{4}^{\prime} & 0 & 0 & 0\\
A_{4}^{\prime} & M_{5}^{\prime} & 0 & 0\\
A_{5}^{\prime} & 0 & M_{5}^{\prime} & 0\\
A_{6}^{\prime} & 0 & 0 & M_{6}^{\prime}
\end{array}\right),\label{eq:Spindof6M2-1}
\end{eqnarray}
with 
\begin{eqnarray}
M_{4}^{\prime} & = & \left(\begin{array}{ccc}
i\omega & -ik & \frac{1}{2}\omega k\\
-ikc_{s}^{2} & i\omega & \frac{1}{2}\omega^{2}\\
\lambda^{\prime}c_{s}^{2}ik+8\lambda\chi_{e}^{0x} & \lambda^{\prime}i\omega & 2D_{b}+\tau_{q}\omega^{2}-i\omega
\end{array}\right),\\
M_{5}^{\prime} & = & \left(\begin{array}{ccc}
i\omega & \frac{1}{2}\omega^{2} & -\frac{1}{2}\omega k\\
\lambda^{\prime}i\omega & 2D_{b}+\tau_{q}\omega^{2}-i\omega & 0\\
2\gamma^{\prime}ik & 0 & -\tau_{\phi}\omega^{2}+i\omega+2D_{s}
\end{array}\right),\label{eq:M5pri}\\
M_{6}^{\prime} & = & -\tau_{\phi}\omega^{2}+i\omega+2D_{s}.
\end{eqnarray}
The off-diagonal matrices $A_{4,5,6}^{\prime}$ are given by
\begin{equation}
A_{4}^{\prime}\equiv\left(\begin{array}{ccc}
0 & 0 & 0\\
8\lambda\chi_{e}^{0y} & 0 & 0\\
8\gamma_{s}\chi_{e}^{xy} & 0 & 0
\end{array}\right),\ A_{5}^{\prime}\equiv\left(\begin{array}{ccc}
0 & 0 & 0\\
8\lambda\chi_{e}^{0z} & 0 & 0\\
8\gamma_{s}\chi_{e}^{xz} & 0 & 0
\end{array}\right),\;A_{6}^{\prime}=\left(\begin{array}{ccc}
8\gamma_{s}\chi_{e}^{yz}, & 0, & 0\end{array}\right).
\end{equation}

The dispersion relations $\omega=\omega(k)$ are derived from 
\begin{equation}
\det\mathcal{M}_{2}^{\prime}=\det M_{4}^{\prime}\cdot(\det M_{5}^{\prime})^{2}\cdot\det M_{6}^{\prime}=0.
\end{equation}
We find that there exist two zero modes coming from the equation $\det M_{5}^{\prime}=0$.
Now, let us focus on the nonzero modes. The $\det M_{6}^{\prime}=0$
gives two non-hydrodynamic modes 
\begin{equation}
\omega=\frac{1}{2\tau_{\phi}}(i\pm\sqrt{8D_{s}\tau_{\phi}-1}).
\end{equation}
From $\det M_{4}^{\prime}=0$ and $\det M_{5}^{\prime}=0$, we obtain
the dispersion relation in small $k$ limit, 
\begin{eqnarray}
\omega & = & \pm c_{s}k\mp2ic_{s}\lambda\chi_{e}^{0x}D_{b}^{-1}k^{2}+O(k^{3}),\label{eq:Sound}\\
\omega & = & \left[i\pm\sqrt{-4D_{b}(2\tau_{q}-\lambda^{\prime})-1}\right](2\tau_{q}-\lambda^{\prime})^{-1}+O(k),\\
\omega & = & \frac{1}{2\tau_{\phi}}(i\pm\sqrt{8D_{s}\tau_{\phi}-1})+O(k),
\end{eqnarray}
and, in large $k$ limit,
\begin{eqnarray}
\omega & = & \pm k\sqrt{\frac{c_{s}^{2}(3\lambda^{\prime}+2\tau_{q})}{2\tau_{q}-\lambda^{\prime}}}+\frac{4i\lambda^{\prime}}{(2\tau_{q}-\lambda^{\prime})(2\tau_{q}+3\lambda^{\prime})}\nonumber \\
 &  & \quad\mp\frac{8\lambda\chi_{e}^{0x}}{c_{s}\sqrt{\left(\lambda^{\prime}-2\tau_{q}\right)\left(3\lambda^{\prime}+2\tau_{q}\right)}}+O(k^{-1}),\\
\omega & = & \frac{i\pm\sqrt{-1-4D_{b}(2\tau_{q}+3\lambda^{\prime})}}{2\tau_{q}+3\lambda^{\prime}}+O(k^{-1}),\\
\omega & = & \pm\sqrt{\frac{2\gamma^{\prime}\tau_{q}}{(2\tau_{q}-\lambda^{\prime})\tau_{\phi}}}k+i\frac{[\tau_{q}(2\tau_{q}-\lambda^{\prime})+\lambda^{\prime}\tau_{\phi}]}{2\tau_{q}\tau_{\phi}(2\tau_{q}-\lambda^{\prime})}+O(k^{-1}),\label{eq:SolM5pri-Bigk-1}\\
\omega & = & \frac{i\pm\sqrt{-1-8D_{b}\tau_{q}}}{2\tau_{q}}+O(k^{-1}).\label{eq:SolM5pri-Bigk-2}
\end{eqnarray}

The causality conditions (\ref{eq:casuality_01}, \ref{eq:causality_02})
require,
\begin{equation}
0\leq\frac{c_{s}^{2}(3\lambda^{\prime}+2\tau_{q})}{2\tau_{q}-\lambda^{\prime}}\leq1,\ 0\leq\frac{2\gamma^{\prime}\tau_{q}}{(2\tau_{q}-\lambda^{\prime})\tau_{\phi}}\leq1,\label{eq:casuality_condtion_q_phi}
\end{equation}
which implies that the relaxation times $\tau_{q},\tau_{\phi}$ cannot
be arbitrarily small. It is consistent with the discussion in Sec.
\ref{subsec:Minimal-causal-theory}.

The stability condition (\ref{eq:Stablity_condition_01}) leads to
\begin{equation}
\tau_{q}>\lambda^{\prime}/2,\ D_{s}>0,\ D_{b}<0,\ \chi_{e}^{0x}=0,\label{eq:Stability-Q}
\end{equation}
where $\chi_{e}^{0x}=0$ comes from the stability of the sound mode
(\ref{eq:Sound}). Although the conditions in Eq. (\ref{eq:Stability-Q})
are derived from the small $k$ and large $k$ limits only, we can
implement the Routh-Hurwitz criterion \citep{Gopal2006ControlSP,Gradshteuin:2007book,Kovtun:2019hdm,Bemfica:2019knx,Hoult:2020eho,Bemfica:2020zjp}
to prove that the conditions (\ref{eq:Stability-Q}) are sufficient
and necessary for stability, i.e., if (\ref{eq:Stability-Q}) are
satisfied, then $\mathrm{Im}\ \omega>0$ for all $k\neq0$. Details
for the proof are given below. 

\subsection{Condition (\ref{eq:Stability-Q}) is sufficient and necessary for
the stability \label{sec:RHwithoutCouple}}

As mentioned, we have derived the stability condition (\ref{eq:Stability-Q})
from the linear modes analysis in small and large $k$ limits only.
Now, we implement the Routh-Hurwitz criterion \citep{Gopal2006ControlSP,Gradshteuin:2007book,Kovtun:2019hdm,Bemfica:2019knx,Hoult:2020eho,Bemfica:2020zjp}
to prove that the condition (\ref{eq:Stability-Q}) guarantees stability
for all real nonzero $k$. 

We only need to prove that the nonzero modes derived from $\det M_{4}^{\prime}=0$
and $\det M_{5}^{\prime}=0$ satisfy $\mathrm{Im}\ \omega>0$ for
all $k$. First, we discuss the modes coming from the $\det M_{4}^{\prime}=0$.
The $\det M_{4}^{\prime}=0$ gives 
\begin{equation}
a_{0}\omega^{4}-ia_{1}\omega^{3}-a_{2}\omega^{2}+ia_{3}\omega+a_{4}=0,\label{eq:EqM4prime}
\end{equation}
with 
\begin{eqnarray}
a_{0} & = & \frac{1}{2}(2\tau_{q}-\lambda^{\prime}),\nonumber \\
a_{1} & = & 1,\nonumber \\
a_{2} & = & \frac{1}{2}c_{s}^{2}k^{2}(3\lambda^{\prime}+2\tau_{q})-2D_{b},\nonumber \\
a_{3} & = & c_{s}^{2}k^{2},\nonumber \\
a_{4} & = & -2c_{s}^{2}D_{b}k^{2}.
\end{eqnarray}
We redefine $\omega=-i\Delta$ and rewrite Eq. (\ref{eq:EqM4prime})
as,
\begin{equation}
a_{0}\Delta^{4}+a_{1}\Delta^{3}+a_{2}\Delta^{2}+a_{3}\Delta+a_{4}=0.
\end{equation}
Notice that the coefficients $a_{0,1,2,3,4}$ are pure real. According
to the Routh-Hurwitz criterion \citep{Gopal2006ControlSP,Gradshteuin:2007book,Kovtun:2019hdm,Bemfica:2019knx,Hoult:2020eho,Bemfica:2020zjp},
the stability condition (\ref{eq:Stablity_condition_01}), i.e., $\mathrm{Im}\ \omega>0$
or $\mathrm{Re}\ \Delta<0$, is fulfilled for all nonzero $k$ if
and only if 
\begin{eqnarray}
a_{i} & > & 0,\nonumber \\
a_{1}a_{2}a_{3}-a_{1}^{2}a_{4}-a_{0}a_{3}^{2} & > & 0.
\end{eqnarray}
When the conditions in Eq. (\ref{eq:Stability-Q}) are fulfilled,
the first inequality $a_{i}>0$ are automatically satisfied. The second
inequality can be expressed as $\lambda^{\prime}=2\lambda/[e_{(0)}+p_{(0)}]>0$,
which has already been guaranteed by entropy principle (\ref{eq:coefficients_01}).
Thus the modes derived from $\det M_{4}^{\prime}=0$ are stable for
all $k$ if condition (\ref{eq:Stability-Q}) is satisfied.

Second, we consider the nonzero modes derived from $\det M_{5}^{\prime}=0$.
The $\det M_{5}^{\prime}=0$ gives $\omega=0$ or 
\begin{equation}
a_{0}^{\prime}\omega^{4}-ia_{1}^{\prime}\omega^{3}-a_{2}^{\prime}\omega^{2}+ia_{3}^{\prime}\omega+a_{4}^{\prime}=0,\label{eq:EqM5prime}
\end{equation}
where 
\begin{eqnarray}
a_{0}^{\prime} & = & \frac{1}{2}\tau_{\phi}(2\tau_{q}-\lambda^{\prime}),\nonumber \\
a_{1}^{\prime} & = & \tau_{\phi}+\frac{1}{2}(2\tau_{q}-\lambda^{\prime}),\nonumber \\
a_{2}^{\prime} & = & 1+D_{s}(2\tau_{q}-\lambda^{\prime})+k^{2}\gamma^{\prime}\tau_{q}-2D_{b}\tau_{\phi},\nonumber \\
a_{3}^{\prime} & = & \gamma^{\prime}k^{2}+2D_{s}-2D_{b},\nonumber \\
a_{4}^{\prime} & = & -4D_{b}D_{s}-2D_{b}\gamma^{\prime}k^{2}.
\end{eqnarray}
Similarly, the Routh-Hurwitz criterion provides the necessary and
sufficient conditions for $\mathrm{Im}\ \omega>0$ in Eq. (\ref{eq:EqM5prime}),
\begin{eqnarray}
a_{i}^{\prime} & > & 0,\\
a_{1}^{\prime}a_{2}^{\prime}a_{3}^{\prime}-a_{1}^{\prime2}a_{4}^{\prime}-a_{0}^{\prime}a_{3}^{\prime2} & > & 0.\label{eq:RH-2}
\end{eqnarray}
Each $a_{i}^{\prime}>0$ does not give new constraints for stability.
We now show that the second inequality holds for all $k$ if the conditions
in Eq. (\ref{eq:Stability-Q}) are fulfilled. Define a new function
$F(D_{b},D_{s},k)$,
\begin{eqnarray}
F(D_{b},D_{s},k) & \equiv & a_{1}^{\prime}a_{2}^{\prime}a_{3}^{\prime}-a_{1}^{\prime2}a_{4}^{\prime}-a_{0}^{\prime}a_{3}^{\prime2}\nonumber \\
 & = & 4\tau_{\phi}^{2}D_{b}^{2}+\frac{1}{2}[8D_{s}(2\tau_{q}-\lambda^{\prime})\tau_{\phi}+G(k)]D_{b}+H(D_{s},k),\label{eq:FDDk}
\end{eqnarray}
with
\begin{eqnarray}
G(k) & \equiv & -(2+k^{2}\gamma^{\prime}\lambda^{\prime})(2\tau_{q}-\lambda^{\prime})-2[2+k^{2}\gamma^{\prime}(3\lambda^{\prime}-4\tau_{q})]\tau_{\phi},\\
H(D_{s},k) & \equiv & \frac{1}{2}(2D_{s}+k^{2}\gamma^{\prime})(2\tau_{q}-\lambda^{\prime})[1+D_{s}(2\tau_{q}-\lambda^{\prime})+k^{2}\gamma^{\prime}\tau_{q}]\nonumber \\
 &  & +\frac{1}{2}(2D_{s}+k^{2}\gamma^{\prime})(2+k^{2}\gamma^{\prime}\lambda^{\prime})\tau_{\phi}.
\end{eqnarray}
Since $\tau_{q}>\lambda^{\prime}/2$ in Eq. (\ref{eq:Stability-Q}),
we have $H(D_{s},k)>0$ for any $k$ and any $D_{s}>0$.

Then, we discuss two cases. When
\begin{equation}
8D_{s}(2\tau_{q}-\lambda^{\prime})\tau_{\phi}+G(k)\leq0,
\end{equation}
we find $F(D_{b},D_{s},k)>0$ for any $D_{b}<0$. In another case,
$8D_{s}(2\tau_{q}-\lambda^{\prime})\tau_{\phi}+G(k)>0$, i.e.,
\begin{equation}
D_{s}>\frac{-G(k)}{8(2\tau_{q}-\lambda^{\prime})\tau_{\phi}},\label{eq:Ds-1}
\end{equation}
for each fixed $D_{s}>0$ and $k$, the function $F(D_{b},D_{s},k)$
gets its minimal value 
\begin{eqnarray}
F(D_{b},D_{s},k) & \geq & F(D_{b},D_{s},k)\big|_{D_{b}=-[8D_{s}(2\tau_{q}-\lambda^{\prime})\tau_{\phi}+G(k)]/(16\tau_{\phi}^{2})}\nonumber \\
 & = & \frac{1}{64\tau_{\phi}^{2}}(2+k^{2}\gamma^{\prime}\lambda^{\prime})(\lambda^{\prime}-2\tau_{q}-2\tau_{\phi})^{2}\nonumber \\
 &  & \ \times[16\tau_{\phi}D_{s}-2-k^{2}\gamma^{\prime}(\lambda^{\prime}-8\tau_{\phi})].\label{eq:FDDkmin}
\end{eqnarray}
 at
\begin{equation}
D_{b}=-[8D_{s}(2\tau_{q}-\lambda^{\prime})\tau_{\phi}+G(k)]/(16\tau_{\phi}^{2}).
\end{equation}
Substituting Eq. (\ref{eq:Ds-1}) into Eq. (\ref{eq:FDDkmin}) leads
to
\begin{eqnarray}
F(D_{b},D_{s},k) & \geq & \frac{(2+k^{2}\gamma^{\prime}\lambda^{\prime})^{2}(\lambda^{\prime}-2\tau_{q}-2\tau_{\phi})^{2}(2\tau_{q}-\lambda^{\prime}+4\tau_{\phi})}{64(2\tau_{q}-\lambda^{\prime})\tau_{\phi}^{2}}>0,
\end{eqnarray}
where we have used $\tau_{q}>\lambda^{\prime}/2$ in Eq. (\ref{eq:Stability-Q}).
Thus, the nonzero modes derived from $\det M_{5}^{\prime}=0$ are
stable for all $k$ if the conditions in Eq. (\ref{eq:Stability-Q})
are fulfilled.

Therefore, the conditions in Eq. (\ref{eq:Stability-Q}) are sufficient
and necessary for the stability of fluids with $q^{\mu}$ and $\phi^{\mu\nu}$
only.

\subsection{Analysis for case II \label{subsec:Analysis-for-II}}

We now consider a more general case where $q^{\mu}$ and $\phi^{\mu\nu}$
are coupled as shown in Eqs.(\ref{eq:Type1q-coupled},\ref{eq:Type1phi-coupled}).
Here we consider the $q^{\mu}$ and $\phi^{\mu\nu}$ only and neglect
other dissipative terms for simplicity. In this case, $M_{5}^{\prime}$
in Eq. (\ref{eq:M5pri}) should be replaced with 
\begin{eqnarray}
M_{5}^{\prime} & = & \left(\begin{array}{ccc}
i\omega & \frac{1}{2}\omega^{2} & -\frac{1}{2}\omega k\\
\lambda^{\prime}i\omega & 2D_{b}+\tau_{q}\omega^{2}-i\omega & g_{1}\omega k\\
2\gamma^{\prime}ik & -\frac{1}{4}g_{2}\omega k & -\tau_{\phi}\omega^{2}+i\omega+2D_{s}
\end{array}\right),\label{eq:M5pri-coupled}
\end{eqnarray}
while the matrix $M_{4}^{\prime}$ is the same as before. The dispersion
relations in Eq. (\ref{eq:SolM5pri-Bigk-1}-\ref{eq:SolM5pri-Bigk-2})
become 
\begin{eqnarray}
\omega & = & \pm\sqrt{\frac{m}{4(2\tau_{q}-\lambda^{\prime})\tau_{\phi}}}k+\frac{1}{2}i\left(\frac{2}{2\tau_{q}-\lambda^{\prime}}+\frac{1}{\tau_{\phi}}-\frac{8\gamma^{\prime}}{m}\right)+\mathcal{O}(k^{-1}),\label{eq:SolM5pri-Bigk-1-coupled}\\
\omega & = & \frac{4\gamma^{\prime}(i\pm\sqrt{-1-D_{b}m\gamma^{\prime-1}})}{m}+\mathcal{O}(k^{-1}),\label{eq:SolM5pri-Bigk-2-coupled}
\end{eqnarray}
where $m=2g_{1}g_{2}+8g_{1}\gamma^{\prime}+g_{2}\lambda^{\prime}+8\gamma^{\prime}\tau_{q}$.
We also notice that the zero modes mentioned before cannot be solved
by introducing the coupling between $q^{\mu}$ and $\phi^{\mu\nu}$. 

Imposing Eq. (\ref{eq:casuality_01}) to the propagating modes in
Eqs.(\ref{eq:Six2Largek-3}-\ref{eq:Six2Largek-4}), the causality
conditions (\ref{eq:casuality_condtion_q_phi}) are replaced with 

\begin{equation}
0\leq\frac{c_{s}^{2}(3\lambda^{\prime}+2\tau_{q})}{2\tau_{q}-\lambda^{\prime}}\leq1,\ 0\leq\frac{m}{4(2\tau_{q}-\lambda^{\prime})\tau_{\phi}}\leq1.\label{eq:casuality_condtion_q_phi-coupled}
\end{equation}
Inserting Eq. (\ref{eq:Stablity_condition_01}) into the new dispersion
relations, we obtain the new stability conditions 
\begin{equation}
\tau_{q}>\lambda^{\prime}/2,\ D_{s}>0,\ D_{b}<0,\ \chi_{e}^{0x}=0,\quad m>8\gamma^{\prime}\left(\frac{2}{2\tau_{q}-\lambda^{\prime}}+\frac{1}{\tau_{\phi}}\right)^{-1}.\label{eq:Stability-Q-coupled}
\end{equation}
Similarly, we can still implement the Routh-Hurwitz criterion to verify
that the conditions in Eq. (\ref{eq:Stability-Q-coupled}) are sufficient
and necessary for stability. 

\subsection{Condition (\ref{eq:Stability-Q-coupled}) is sufficient and necessary
for the stability \label{subsec:prove-case-II}}

Let us now prove that the condition (\ref{eq:Stability-Q-coupled})
ensures $\mathrm{Im}\ \omega>0$ for all nonzero real $k$. Consider
the nonzero modes derived from $\det M_{5}^{\prime}=0$. The $\det M_{5}^{\prime}=0$
gives 
\begin{equation}
a_{0}^{\prime}\omega^{4}-ia_{1}^{\prime}\omega^{3}-a_{2}^{\prime}\omega^{2}+ia_{3}^{\prime}\omega+a_{4}^{\prime}=0,\label{eq:EqM5prime-coupled}
\end{equation}
where 
\begin{eqnarray}
a_{0}^{\prime} & = & \frac{1}{2}\tau_{\phi}(2\tau_{q}-\lambda^{\prime}),\nonumber \\
a_{1}^{\prime} & = & \tau_{\phi}+\frac{1}{2}(2\tau_{q}-\lambda^{\prime}),\nonumber \\
a_{2}^{\prime} & = & 1+D_{s}(2\tau_{q}-\lambda^{\prime})+\frac{1}{8}k^{2}m-2D_{b}\tau_{\phi},\nonumber \\
a_{3}^{\prime} & = & \gamma^{\prime}k^{2}+2D_{s}-2D_{b},\nonumber \\
a_{4}^{\prime} & = & -4D_{b}D_{s}-2D_{b}\gamma^{\prime}k^{2}.
\end{eqnarray}
The necessary and sufficient conditions for $\textrm{Im }\omega>0$
in Eq. (\ref{eq:EqM5prime-coupled}) are
\begin{eqnarray}
a_{i}^{\prime} & > & 0,\\
a_{1}^{\prime}a_{2}^{\prime}a_{3}^{\prime}-a_{1}^{\prime2}a_{4}^{\prime}-a_{0}^{\prime}a_{3}^{\prime2} & > & 0.\label{eq:RH-2-1}
\end{eqnarray}
The first conditions are automatically satisfied when we have the
constraints for stability. Then we need to analyze whether Eq. (\ref{eq:RH-2-1})
is satisfied under the existing constraints.

Define a function $F(D_{b},D_{s},k)$,
\begin{eqnarray}
F(D_{b},D_{s},k) & \equiv & a_{1}^{\prime}a_{2}^{\prime}a_{3}^{\prime}-a_{1}^{\prime2}a_{4}^{\prime}-a_{0}^{\prime}a_{3}^{\prime2}\nonumber \\
 & = & F_{a}D_{b}^{2}+F_{b}D_{b}+F_{c},\label{eq:FDDk-coupled}
\end{eqnarray}
where
\begin{eqnarray}
F_{a} & \equiv & 4\tau_{\phi}^{2},\nonumber \\
F_{b} & \equiv & \left[\frac{1}{2}k^{2}\gamma^{\prime}(2\tau_{q}-\lambda^{\prime})+(4D_{s}+3k^{2}\gamma^{\prime})\tau_{\phi}\right](2\tau_{q}-\lambda^{\prime})-\frac{1}{8}(mk^{2}+8)(2\tau_{\phi}+2\tau_{q}-\lambda^{\prime}),\nonumber \\
F_{c} & \equiv & \frac{1}{16}(2D_{s}+k^{2}\gamma^{\prime})\left\{ 8D_{s}(2\tau_{q}-\lambda^{\prime})^{2}+(2\tau_{q}-\lambda^{\prime})[8+k^{2}(m-8\gamma^{\prime}\tau_{\phi})]+2(8+k^{2}m)\tau_{\phi}\right\} \nonumber \\
 & > & \frac{1}{2}(2D_{s}+k^{2}\gamma^{\prime})\{2\tau_{\phi}+(2\tau_{q}-\lambda^{\prime})[1+D_{s}(2\tau_{q}-\lambda^{\prime})]\}>0.
\end{eqnarray}
When $F_{b}<0$, i.e.,
\begin{eqnarray}
D_{s} & < & \frac{(mk^{2}+8)(2\tau_{\phi}+2\tau_{q}-\lambda^{\prime})}{32(2\tau_{q}-\lambda^{\prime})\tau_{\phi}}-\frac{k^{2}\gamma^{\prime}(2\tau_{q}-\lambda^{\prime})}{8\tau_{\phi}}-\frac{3}{4}k^{2}\gamma^{\prime},
\end{eqnarray}
we get
\begin{equation}
F(D_{b},D_{s},k)>F(0,D_{s},k)=F_{c}>0.
\end{equation}
In another case, $F_{b}\geq0$, i.e.,
\begin{eqnarray}
D_{s} & \geq & \frac{(mk^{2}+8)(2\tau_{\phi}+2\tau_{q}-\lambda^{\prime})}{32(2\tau_{q}-\lambda^{\prime})\tau_{\phi}}-\frac{k^{2}\gamma^{\prime}(2\tau_{q}-\lambda^{\prime})}{8\tau_{\phi}}-\frac{3}{4}k^{2}\gamma^{\prime},
\end{eqnarray}
the function has its minimal value
\begin{eqnarray}
F(D_{b},D_{s},k)_{\textrm{min}} & = & F(D_{b},D_{s},k)|_{D_{b}=-F_{b}/(2F_{a})}\nonumber \\
 & = & -\frac{(2\tau_{\phi}+2\tau_{q}-\lambda^{\prime})^{2}}{1024\tau_{\phi}^{2}}\left\{ 8+k^{2}\left[m-4\gamma^{\prime}\left(2\tau_{q}-\lambda^{\prime}\right)\right]\right\} \nonumber \\
 &  & \times\left\{ 8+k^{2}\left[m-4\gamma^{\prime}\left(2\tau_{q}-\lambda^{\prime}\right)\right]-32k^{2}\tau_{\phi}(\gamma^{\prime}+2D_{s})\right\} \nonumber \\
 & \geq & \frac{\left\{ 8+k^{2}\left[m-4\gamma^{\prime}\left(2\tau_{q}-\lambda^{\prime}\right)\right]\right\} ^{2}\left(2\tau_{\phi}+2\tau_{q}-\lambda^{\prime}\right)^{3}}{1024\tau_{\phi}^{2}(2\tau_{q}-\lambda^{\prime})}>0,
\end{eqnarray}
at
\begin{eqnarray}
D_{b} & = & -\frac{F_{b}}{2F_{a}},\\
D_{s} & = & \frac{(mk^{2}+8)(2\tau_{\phi}+2\tau_{q}-\lambda^{\prime})}{32(2\tau_{q}-\lambda^{\prime})\tau_{\phi}}-\frac{k^{2}\gamma^{\prime}(2\tau_{q}-\lambda^{\prime})}{8\tau_{\phi}}-\frac{3}{4}k^{2}\gamma^{\prime}.
\end{eqnarray}
Therefore, the nonzero modes are stable for all $k$ if the stability
condition (\ref{eq:Stability-Q-coupled}) is satisfied.

\section{Discussions on the stability conditions (\ref{eq:SforMSix_02}) \label{sec:Discussions-about-}}

Here, we discuss the stability conditions (\ref{eq:SforMSix_02}),
i.e., $D_{s}>0$, $D_{b}<0$.

Let us consider an isotropic fluid at equilibrium, i.e., we assume
that there are not preferred directions induced by spin and external
fields. In this case, the variation of spin chemical potential is
\begin{equation}
\delta\omega^{\mu\nu}=\chi^{\mu\nu\alpha\beta}\delta S_{\alpha\beta}+\chi_{e}^{\mu\nu}\delta e,\label{eq:IsotropicLimit}
\end{equation}
with a rank-$4$ tensor $\chi^{\mu\nu\alpha\beta}$ and rank-$2$
tensor $\chi_{e}^{\mu\nu}$. We find that $\chi^{\mu\nu\alpha\beta}$
satisfies $\chi^{\mu\nu\alpha\beta}=-\chi^{\nu\mu\alpha\beta}=-\chi^{\mu\nu\beta\alpha}$.

In an irrotational isotropic background fluid without any external
fields, any rank-$n$ tensor can only be constructed by $u^{\mu},g^{\mu\nu},\partial^{\mu},\epsilon^{\mu\nu\alpha\beta}$.
Back to rank-$4$ tensor $\chi^{\mu\nu\alpha\beta}$, in the linear
modes analysis, we do not need to consider the part in $\chi^{\mu\nu\alpha\beta}$
proportional to space-time derivatives $\partial^{\mu}$ since those
terms in $\chi^{\mu\nu\alpha\beta}\delta S_{\alpha\beta}$ becomes
nonlinear and will be dropped. While the tensor $\epsilon^{\mu\nu\alpha\beta}$
violates the reflection symmetry and cannot be used there. According
to the anti-symmetric properties of $\chi^{\mu\nu\alpha\beta}$, the
only possible expression is 
\begin{equation}
\chi^{\mu\nu\alpha\beta}=\frac{\chi_{1}}{2}(g^{\mu\alpha}g^{\nu\beta}-g^{\mu\beta}g^{\nu\alpha})+\frac{\chi_{2}}{2}(\Delta^{\mu\alpha}\Delta^{\nu\beta}-\Delta^{\mu\beta}\Delta^{\nu\alpha}),\label{eq:Isochi}
\end{equation}
where $\chi_{1}$ and $\chi_{2}$ are scalars.

Substituting Eq. (\ref{eq:Isochi}) into Eq. (\ref{eq:IsotropicLimit}),
we obtain 
\begin{equation}
\delta\omega^{\mu\nu}=\chi_{1}\delta S^{\mu\nu}+\chi_{2}\Delta^{\mu\alpha}\Delta^{\nu\beta}\delta S_{\alpha\beta}.
\end{equation}
One can also write it as 
\begin{eqnarray}
u_{\mu}\delta\omega^{\mu\nu} & = & \chi_{1}u_{\mu}\delta S^{\mu\nu},\\
\Delta^{\mu\alpha}\Delta^{\nu\beta}\delta\omega_{\alpha\beta} & = & (\chi_{1}+\chi_{2})\Delta^{\mu\alpha}\Delta^{\nu\beta}\delta S_{\alpha\beta}.
\end{eqnarray}
From the definitions in Eqs.(\ref{eq:Notations0},\ref{eq:Notations1}),
we then have 
\begin{equation}
D_{s}=4\gamma_{s}(\chi_{1}+\chi_{2}),\ D_{b}=4\lambda\chi_{1}.
\end{equation}
Since $\gamma_{s}>0,\lambda>0$, the stability condition (\ref{eq:SforMSix_02}),
$D_{s}>0,D_{b}<0$, is equivalent to 
\begin{equation}
\chi_{2}>-\chi_{1}>0.\label{eq:chi1chi2}
\end{equation}

The equation of state used in our previous works \citep{Wang:2021ngp,Wang:2021wqq}
corresponds to $\chi_{2}=0$ (see Eq. (17) of Ref. \citep{Wang:2021ngp})
and Eq. (38) of Ref. \citep{Wang:2021wqq}). In that case, Eq. (\ref{eq:chi1chi2})
cannot be satisfied and there exist unstable modes, although the analytic
solutions in Refs. \citep{Wang:2021ngp,Wang:2021wqq} do not rely
on it. For general cases where $\chi_{2}\neq0$, whether the stability
condition (\ref{eq:SforMSix_02}) $D_{s}>0,D_{b}<0$ is satisfied
depends on $\chi_{1},\chi_{2}$, which relates with the equation of
state for $S^{\mu\nu}$ and $\omega^{\mu\nu}$. To determine the value
of $\chi_{1},\chi_{2}$, further investigations should be done from
the microscopic theory.

\bibliographystyle{h-physrev}
\bibliography{qkt-ref20230407}

\end{document}